\documentclass{article}
\usepackage{graphicx}

\usepackage{algorithm}
\usepackage{algorithmic}
\usepackage[bb=dsserif]{mathalpha}
\usepackage{booktabs}
\usepackage{caption}
\usepackage{graphicx} 
\usepackage[round]{natbib}
\usepackage{amsmath}
\usepackage{multirow}
\usepackage[margin=1in]{geometry}

\renewcommand{\vec}[1]{\ensuremath{#1}}

\newcommand{\newlinecell}[2][c]{%
  \begin{tabular}[#1]{@{}c@{}}#2\end{tabular}}

\begin{document}
\title{Fast and Accurate Node-Age Estimation Under Fossil Calibration Uncertainty Using the Adjusted Pairwise Likelihood}

\author{Greg M. Ellison$^{1}$, Liang Liu$^{1, \ast}$,
\textit{$^{1}$~Department of Statistics, University of Georgia, Athens, 30601, USA}
\\
\\[2pt]
\textit{*Correspondance should be sent to LL. Email: lliu@uga.edu}}
\markboth%
{Ellison \& Liu}
{Node Dating with the Pairwise Likelihood}

\maketitle
\abstract{Estimating divergence times from molecular sequence data is central
to reconstructing the evolutionary history of lineages. Although Bayesian
relaxed-clock methods provide a principled framework for incorporating fossil
information, their dependence on repeated evaluations of the full phylogenetic
likelihood makes them computationally demanding for large genomic datasets.
Furthermore, because disagreements in divergence-time estimates often arise
from uncertainty or error in fossil placement and prior specification, there is
a need for methods that are both computationally efficient and robust to
fossil-calibration uncertainty. In this study, we introduce fast and accurate
alternatives based on the phylogenetic pairwise composite likelihood,
presenting two adjusted pairwise likelihood (APW) formulations that employ
asymptotic moment-matching weights to better approximate the behavior of the
full likelihood within a Bayesian MCMC framework. Extensive simulations across
diverse fossil-calibration scenarios show that APW methods produce node-age
estimates comparable to those obtained from the full likelihood while offering
greater robustness to fossil misplacement and prior misspecification, due to
the reduced sensitivity of composite likelihoods to local calibration errors.
Applied to a genome-scale dataset of modern birds, APW methods recover
divergence time patterns consistent with recent studies, while reducing
computational cost by more than an order of magnitude. Overall, our results
demonstrate that adjusted pairwise likelihoods provide a calibration-robust and
computationally efficient framework for Bayesian node dating, especially suited
for large phylogenomic datasets and analyses in which fossil priors may be
uncertain or imperfectly placed.  }
\newpage



\renewcommand{\algorithmicrequire}{\textbf{Input:}}
\renewcommand{\algorithmicensure}{\textbf{Output:}}

\renewcommand*{\vec}[1]{\boldsymbol#1}


\newcommand{\sumkl}[1]{\ensuremath{\sum_{k=1}^{#1-1} \sum_{l=k+1}^{#1}}}
\newcommand{\prodkl}[1]{\ensuremath{\prod_{k=1}^{#1-1} \prod_{l=k+1}^{#1}}}
\newcommand{\sumi}[1]{\ensuremath{\sum_{i=1}^{#1}}}
\newcommand{\prodi}[1]{\ensuremath{\prod_{i=1}^{#1}}}


\newcommand{\trace}{\ensuremath{\text{trace}}}
\newcommand{\argmaxu}[1]{\underset{#1}{\text{argmax}}~}
\newcommand{\argminu}[1]{\underset{#1}{\argmin}~}
\renewcommand{\algorithmicrequire}{\textbf{Input:}}
\renewcommand{\algorithmicensure}{\textbf{Output:}}
\newcommand{\lc}{\ensuremath{L_c(\psi; \delta)}}
\newcommand{\llc}{\ensuremath{\ell_{c}(\psi| \delta, X )}}
\newcommand{\lfull}{\ensuremath{L(\theta)}}
\newcommand{\llfull}{\ensuremath{\ell(\theta)}}

\newcommand{\prior}[1]{\ensuremath{\pi(#1)}}
\newcommand{\postc}{\ensuremath{\pi_c( \psi | X, \delta)}}
\newcommand{\postfull}{\ensuremath{\pi(\theta ; X)}}
\newcommand{\postca}{\ensuremath{\pi^{adj}_c( \psi ; X, \delta)}}

\newcommand{\convdist}{\ensuremath{\overset{d}{\to}}}

\newcommand{\mcle}[1]{\ensuremath{\hat{#1}_c}}
\newcommand{\cscore}[2]{\ensuremath{u_c(#1)}}
\newcommand{\cJ}[2]{\ensuremath{H_c(#1)}}
\newcommand{\cH}[2]{\ensuremath{J_c(#1)}}
\newcommand{\score}[1]{\ensuremath{u_c(#1)}}


\section{Introduction}

Estimating divergence times, often referred to as node-dating, is a foundational
task in evolutionary biology. It allows researchers to place evolutionary events
on an absolute timescale, transforming phylogenetic trees from representations
of relationships into historical narratives of life on Earth \citep{Tamura_2012,
Kumar_2022}. The molecular clock model \citep{Zuckerkandl1965, Mulvey2025}
provides a link between sequence evolution and evolution of species on a real
time-scale, and carbon dated fossils can be used to calibrate the evolutionary
rate of the molecular clock. Although the likelihood function relating the
phylogenetic model to the alignment data is expressed in terms of
\textit{evolutionary distances} (normalized lengths in units of expected number
of mutations per site), information relating mutation rates to real-world time
units translates the height of nodes on the tree to real-world time units.
Relaxed clock models allow the mutation rates of molecular characters to vary
across the tree, for example by assuming the mutation rates at each branch of
the tree are independently distributed according to some prior distribution
\citep{Drummond2006, Lepage2007}. Since the node ages and mutation rates are
confounded in the likelihood function, accurate estimation of node ages relies
on accurate time calibrations using fossil evidence \citep{DosReis2012}.
However, uncertainty about the ages of fossils can introduce considerable
uncertainty in estimated divergence times \citep{Rannala_2007, DosReis2012}.
Much attention has been given to the role of fossil calibrations in node dating
clock analyses, particularly the fact that node age estimates are highly
dependent on fossil calibrations \citep{Yang2005, Heads2005, Hug2007,
Forest2009, DosReis2012, Luo2019, Nguyen2020}. Critically, according to the
\textit{infinite-sites} theory of \citet{Rannala_2007}, uncertainties in the
estimates of divergence times do not resolve with more sequence data, and the
quality of node age estimates is highly dependent on the quality of fossil
information used to calibrate the node ages.

Bayesian methods are commonly used for node-dating due to their ability to
incorporate prior information in the form of fossils calibrations. One or more fossils
are placed at nodes of the tree topology to constrain the topology according to
the ancestral relationship of the fossil to sampled taxa and a prior
distribution is placed on the node age according to evidence the fossil age
\citep{Parham2011}. In this way, a prior distribution is imposed on the ages of
all nodes, and the prior information is used to disentangle the divergence times
and mutation rates \citep{Inoue2009,Yang2014,Nascimento2017}. Inference about
node ages is made by examining posterior credible sets of node ages samples from
the MCMC chain; the tree topology is often assumed fixed, so no problems arise
from needing to consider nodes only present in some sampled tree topologies,
although with MCMC sampling it is possible to integrate out the uncertainty due
to the unknown topology by sampling from the tree topology posterior
distribution. A well-known limitation of the fully Bayesian approach is the
substantial computational burden created by repeated likelihood evaluations
along long MCMC chains. MCMCTree addresses this issue by employing a likelihood
approximation to accelerate computation \citep{Rannala_2007, Reis2011}, yet the
method remains computationally demanding for datasets with long sequences, a
difficulty shared by other widely used programs such as MrBayes (reference) and
BEAST2 (reference). The challenge has become even more pronounced with the
proliferation of high-throughput sequencing technologies, which produce
large-scale genomic datasets for phylogenetic inference \citep{Liu2015,
Patane2017, Young2019}. Although likelihood-based approaches, including both
maximum likelihood and Bayesian inference, remain the statistical gold standard
due to their desirable theoretical properties and principled frameworks for
model selection and parameter estimation, their computational requirements can
be prohibitive for large phylogenies with many taxa or extensive sequence data.

\textit{Composite likelihoods} have become popular for their ease of
construction and computation \citep{Varin2008, Varin2011}. The pairwise composite
likelihood, for example, constructs an approximate likelihood by combining 
marginal likelihoods of pairs of data; in the case of a DNA sequence
alignment of \(M\) taxa, this likelihood is computed by considering only pairs
of nucleotides rather than the full site patterns of \(M\) characters. This
approach can result in a considerable computational advantage, although it has
theoretical drawbacks due to ignoring the state of ancestral sequences and the
dependency among sequences \citep{Farris1985, Steel2009}. However, methods have
been developed to mitigate the drawbacks that arise when using the composite likelihood
as a misspecified likelihood function \citep{Varin2008, Varin2011}, including in
the Bayesian setting \citep{Pauli2011, Ribatet2012, Shaby2014, Miller2021}; as the 
marginal likelihoods that comprise the pairwise likelihood are valid likelihoods themselves,
the composite pairwise likelihood retains good estimation properties. Adjustment
weights based on asymptotic properties of the composite Likelihood Ratio Test
(LRT) statistic are often used to correct the composite likelihood, and result in
improved parameter estimation relative to the unadjusted composite likelihood
\citep{Pauli2011, Ribatet2012, Cattelan2015}. The pairwise likelihood has been
studied previously for phylogeny inference \citep{Holder2011, Wu2009, Wu2010},
and recent works including \citet{Peng2022} and \citet{Kong2024} have applied
composite likelihood methods to tree estimation under the multi-species
coalescent and use of phylogenetic networks to efficiently estimate tree
topologies in the presence of hybridization, respectively. \citet{Liu2010} 
was the first paper to use the composite likelihood to estimate species
trees under the mult-species coalescent model.

For large sequence datasets, we propose a Bayesian divergence-time framework
that replaces the full phylogenetic likelihood with a pairwise composite
likelihood. Since the pairwise likelihood involves a number of terms
proportional to \(c^{2} {M \choose 2}\), where \(M\) denotes the number of taxa
and \(c\) the number of character states in the substitution model, its
computational cost does not scale with sequence length. In contrast, the full
phylogenetic likelihood becomes increasingly expensive as alignments grow
longer. The pairwise composite likelihood therefore offers a computationally
efficient alternative for analyses based on long sequence alignments. In this
study, we introduce adjusted pairwise (APW) likelihoods and demonstrate that
they enable fast and accurate estimation of divergence times. We assess the
performance of the APW approach using a comprehensive simulation study modeled
on the node-age analysis of modern birds by \citet{Wu2024}. Although the avian
tree of life has been examined extensively over the past two decades
\citep{Hackett2008, Jarvis2014, Prum2015, Wu2024}, the divergence between modern
birds (\textit{Neoaves}) and flightless birds (\textit{Paleognathae}) remains
contentious, due in part to the placement and uncertainty of fossil calibrations
\citep{Claramunt2024, Wu2024a}. Across a wide range of fossil-calibration
scenarios, our simulations show that APW methods produce node-age estimates
comparable to those obtained from the full likelihood while exhibiting greater
robustness to fossil misplacement and prior misspecification. This robustness
arises from the reduced sensitivity of composite likelihoods to local
calibration errors. These results demonstrate that APW likelihoods provide a
calibration-resilient and computationally efficient framework for Bayesian node
dating.

\section{Background} 

\subsection{Composite likelihoods}

Suppose data \(X = \{ X_i, i=1,\dots,n \},\) are independently generated from
the probability distribution \(f\) parameterized by the vector \(\theta \in
\Theta\). Further suppose \( \theta = (\phi, \gamma) \), where  \(\phi \) is a \(p\)
dimensional vector of parameters of interest and \(\gamma \) is a \(k\)
dimensional vector of nuisance parameters. The likelihood of \(\theta\) is the
joint distribution of the data, \(f(X | \theta)\), viewed as a function of the
parameters and is referred to as the \textit{full} likelihood function. The
\textit{composite} likelihood is an approximate likelihood function constructed
by treating the marginal likelihoods of subsets of the data as independent
\citep{Lindsay_1988, Varin2011}. For marginal events \((A_s), s=1,\dots,S\), let
\(f_s(\theta | X) \propto f(\theta | X \in A_s )\) be the marginal likelihood of
\(A_s\). Then the composite marginal likelihood is defined as \(f_c = \prod_s
f_s(\theta | X)\). Under typical regularity conditions, a central limit theorem
ensures the composite likelihood maximum likelihood estimator is asymptotically
normal, with covariance matrix \(G(\theta)^{-1}\), where \(G\) is defined below 
\citep{Varin2011}.
In the phylogenetic setting, \citep{Holder2011} show that the maximum pairwise likelihood 
estimates are consistent for the tree topology and branch lengths. 

In the Bayesian setting, the parameters are viewed as random, and inference is made 
by sampling parameter values from the posterior distribution of the parameters 
given the observed data. 
Given prior \(\pi(\theta) = \pi(\phi)\pi(\gamma)\) (assumed to be a proper prior),
we denote the typical posterior distribution \(\pi(\theta|X) \propto f_c(\theta | X)
\pi(\theta) \) and as the composite posterior distribution 
\begin{equation}  
    \pi_c(\theta | X) = \frac{f_c(\theta | X) \pi(\theta)}{\int_{\Theta} f_c(\theta | X) \pi(\theta) d\theta}. 
    \label{eqn:pw_posterior}
\end{equation}

The construction of the composite likelihood treats the marginal likelihoods as
independent components, ignoring any higher-order dependence structure captured
by the full likelihood function.  Thus, there is a loss of information due to
use of the composite likelihood relative to the full likelihood, and credible
sets obtained from posterior distributions based on composite likelihoods may
not provide reasonable coverage probabilities of parameters of interest.
\citet{Pauli2011} and \citet{Ribatet2012} apply composite likelihoods to
Bayesian inference of spatial extreme models, and adjust the resulting
posteriors to recover approximate frequentist asymptotic properties of the
composite likelihood. \citet{Shaby2014} finds a similar adjustment with exact
frequentist coverage properties, and \citet{Wu2017} derive a Bernstein-Von Mises
theorem from the composite posterior.  \citet{Miller2021} unites various similar
approaches with the idea with the notion of \textit{Generalized Posteriors}, and
formalizes regularity conditions for generalized posterior concentration,
asymptotic normality, and frequentist coverage set properties. Inference for
\(\theta\) based on the composite likelihood, denoted \(f_c(\theta | X)\) is
often justified via asymptotic results of likelihood ratio test statistics
\citep{Rotnitzky1990, Chandler2007}. Strategies to adjust the composite
likelihood typically make use of the asymptotic distribution of the composite
likelihood ratio test (LRT) statistic \citep{Geys1999,Varin2008,Chandler2007} to
derive a weight for the composite likelihood; we discuss the weights in more
detail below, and for now define an adjusted pairwise likelihood (APW) as \( f_c^a(\theta | X) =
f_c(\theta | X)^w \) for some adjustment weight \(w > 0\), with corresponding
log-likelihood \(\ell_c^a = w \ln f_c(\theta | X). \)

To discuss the asymptotic properties of estimates based on the composite
posterior, we denote the `true' values of the parameters \(\theta_0 = (\phi_0,
\gamma_0)\).  We also denote the maximum composite likelihood estimator
\(\hat{\theta}_c\), defined as 
\(\hat{\theta}_c = \argmaxu{\theta}
\ell_c(\theta) \), where 
\( \ell_c(\theta) \) is the logarithm of the composite likelihood.  
We define the composite score function \( u_c(\theta) =
\nabla_{\theta} \ell_c(\theta) \), and matrices \(J(\theta) = E_{X}[
u_c(\theta)' u_c(\theta) ]\) and  \(H(\theta) = -E_{X}[\nabla_{\theta}
u_c(\theta) ]\), respectively referred to as the \textit{sensitivity} and
\textit{variability} matrices. Since the estimate is from a linear combination
of the data margins and not the full likelihood, \(J(\theta_0) \neq H(\theta_0)
\), and composite likelihood asymptotic theory requires the \textit{Godambe
Information Matrix}
\(G(\theta_0) = H(\theta_0) J^{-1}(\theta_0) H(\theta_0)\). 
If interest is on the parameters \(\phi\), partition the matrices \[
    H(\theta_0) = \left[ \begin{array}{cc} H_{\phi\phi} & H_{\phi\gamma} \\
            H_{\gamma\phi} & H_{\gamma\gamma} \end{array}
            \right]~\text{and}~ H(\theta_0)^{-1} = \left[ \begin{array}{cc}
            H^{\phi\phi} & H^{\phi\gamma} \\ H^{\gamma\phi} &
    H^{\gamma\gamma} \end{array}  \right], \] 
where \(H_{\gamma}\) refers to the block of the partitioned matrix \(M\)
corresponding to the parameter \(\gamma\). For testing parameter value
\(\theta_0\) the composite LRT statistic is defined as \( \Lambda_c = -2
(\ell_c(\hat\theta_c) - \ell_c(\theta_0))\); which has asymptotic distribution
\( \sum_{i=1}^p \lambda_i Z_i^2\), where \( Z_i^2 \) are independent chi-square
random variables and the \(\lambda_i\) are the eigenvalues of \(
(H^{\phi\phi})^{-1} G^{\phi\phi}\), calculated under the null hypothesis. Here,
\(G^{\phi\phi}\) is the submatrix of the inverse of \(G(\theta_0)\) which
corresponds to the parameter \(\phi\). In the case that \(\theta = \phi\), this
simplifies to \(H^{-1}(\theta_0) J(\theta_0)\); from now on we consider only
this case. We consider two so-called \textit{magnitude} 
adjustment weights as described in \citet{Ribatet2012}: 
denoting by \(\lambda_i\) the eigenvalues of \( H^{-1}(\theta_0)
J(\theta_0)\), the first is \( w_1 = p / (\sum_{i=1}^p \lambda_i)\),
such that \(E[w_1 \Lambda_c] = E[\Lambda]\)
\citep{Rotnitzky1990}. The second proposed weight matches the first two moments
is \( w_2 = \sum_{i=1}^p \lambda_i / \sum_{i=1}^p
\lambda_i^2\), such that the composite LRT converges in
distribution to \(\chi^2_{\nu}\), where \(\nu = (\sum_{i=1}^p{\lambda_i})^2 /
\sum_{i=1}^p{\lambda_i^2}\) \citep{Varin2008}. For 
more possible approaches see, for example, \citet{Ribatet2012} or \citet{Shaby2014}. 
In what follows, we denote the adjusted pairwise
likelihoods based on \(w_1\) and \(w_2\) as APW1 and APW2 respectively.

\section{Materials and Methods}

\newcommand{\hpii}[2]{\ensuremath{p_{#1}(\hat{\tau}^c_{#2})}}
\newcommand{\hppii}[2]{\ensuremath{p^{(1)}_{#1}(\hat{\tau}^c_{#2})}}
\newcommand{\hpppii}[2]{\ensuremath{p^{(2)}_{#2}(\hat{\tau}^c_{#2})}}

\newcommand{\pd}[2]{\frac{\partial{#1}}{\partial{#2}}}
\newcommand{\pdd}[2]{\frac{\partial^2{#1}}{\partial{#2}^2}}
\newcommand{\pdt}[3]{\frac{\partial^2{#1}}{\partial{#2}\partial{#3}}}

\newcommand{\pii}[2]{\ensuremath{p_{#1}^{#2}}}
\newcommand{\ppii}[2]{\ensuremath{p_{#1}^{#2}}}
\newcommand{\pppii}[2]{\ensuremath{(p_{#1}^{#2})^{(2)}}}

\subsection{Adjusted pairwise likelihood for phylogenetic inference}

Given a DNA sequence alignment \(X\) consisting of \(N\) sites and \(M\) taxa,
the goal is to estimate the ages, \(y = (y_j), j=1, \dots, y_{M-1} \), of the
internal nodes of the tree. We assume a rooted, bifurcated tree topology \(T\)
is fixed and known. We adopt a \textit{branch-wise} relaxed molecular clock, in
which the mutation rate varies across the tree, but is assumed to be constant
along each branch \citep{Lepage2007}; we define the model in terms of the branch
lengths in units of millions of years, \(b = (b_i), i=1, \dots, 2M-2\) 
and overall mutation rate scalar \(\mu\).

We assume a substitution model with parameters denoted \(\eta\). We will
consider the \(JC\) \citep{Jukes1969} and \(GTR\) \citep{Tavare1986}
substitution models, as the \(GTR\) model is likely the most commonly employed
substitution model in practice, and the JC model is the simplest 
claim above special case of the \(GTR\) model, in which all relative
substitution rates are identical and all nucleotide frequencies are identical.
For simplicity in computing the pairwise likelihood weight, we assume the
substitution model parameters are fixed and known; in practice, they may be
estimated directly from the data and fixed prior to employing the pairwise
likelihood adjustment weight. Assuming the substitution model parameters are
fixed allows computing the eigenvalues of the simpler matrix \( H^{-1} J \)
defined above for computing the pairwise likelihood adjustment weight. We assume
the site rates follow the Discrete Gamma distribution of \citep{Yang1994} with
\(C\) rate categories and shape parameter \(\alpha\); we denote the discrete
site substitution rates as \((\kappa_i), i=1, \dots, C \). Because \(\alpha\) is
not identifiable from the pairwise likelihood for many cases of the \(GTR\)
substitution model (including the \(JC\) model ) \citep{Wu2010}, we consider
\(\alpha\) to be fixed and known when computing the pairwise likelihood.

The composite posterior can be written as 
\[ f_c(y, \mu, \tau, b, \eta, \delta | X) = \frac{
    f_c(X | \tau(\mu, b), \eta) f(b|y, \sigma) f(y | \delta)
    \pi(\tau) \pi( b) \pi(\mu) \pi( y) \pi(\eta) }{m_c(X)} ,
\]
where the pairwise likelihood, defined below, is written as \(f_c(X | \tau(\mu,  b))\)
since the normalized branch lengths are computed as a function of the mutation
rate \(\mu\) and branch lengths \(b\). For convenience, from now on the pairwise
likelihood will simply be denoted \(f_c(X | \tau, \eta) \). Here, \(f(y | \delta)\) 
is the prior distribution on node ages and \(f(b | y, \sigma)\) is the 
prior distribution on branch lengths $b$. We employ a birth-death process 
and independent lognormal branches \citep{Yang2005, Lepage2007}, and \(\delta\) 
and \(\sigma\) represent the hyperparameters associated with these priors.  
For a more detailed treatment, see \citep{Lepage2007}. The denominator
\(m_c(X)\) is the marginal data likelihood under the composite likelihood, and
is obtained by integrating and summing over the joint parameter space. Inference
is made by sampling from the posterior using Markov Chain Monte Carlo (MCMC)
methods \citep{Yang2005, Yang_2014}. 

The alignment matrix is \(X = (X_1, \dots, X_M)\), and the full joint
distribution of the data given the tree topology and branch lengths is denoted
\(f(X | \tau , \eta)\). The elements of the alignment matrix \(X\) are elements
of the character set \(\mathcal{X} = (A,C,G,T)\), and we denote by \(
d=1,\dots,16\) the sixteen doublets made up of pairs of characters in \(\mathcal{X}\).
The composite pairwise likelihood is defined as \(f_c(\tau | X) =
\prod_{k=1}^{M-1} \prod_{l=k+1}^{M} f(X_k, X_l | \tau)\), and the log composite
pairwise likelihood is \(\ell_c = \ln f_c \).  For taxa \(k\) define
\(\mathcal{V}(k,l)\) as the path through the root connecting taxa \(X_k\) and
\(X_l\). The distance separating sequences \(X_k\) and \(X_l\) is \(\nu_{kl} =
\sum_{i \in \mathcal{V}(k,l)} \tau_i\). We also define the set of pairs
connected by branch \(\tau_i\) as \(\mathcal{D}_{\tau_i} = \{(k,l) : \tau_i \in
\mathcal{V}(k,l)\}\).  The observed count of doublet \(d\) in pair \(k, l\) is
denoted \(n^{kl}_{d}\), and the adjusted pairwise log-likelihood of alignment \(X\) is
obtained as 
\begin{align*} \ell_c(\tau) &= w \sumkl{M} \sum_{d=1}^{D} 
    n^{kl}_{d} \log \left( p^{kl}_{d} \right),\\
    \label{eqn:pw_ll}
\end{align*}
where \(p^{kl}_{d}\) represents the probability of doublet \(d\) in sequences
separated by distance \(\nu_{kl}\) and \(w\) is the adjustment weight. The
weights for the adjusted pairwise likelihoods APW1 and APW2 are defined,
respectively, as \( w_1 = p / (\sum_{i=1}^p \lambda_i)\) and \( w_2 =
\sum_{i=1}^p \lambda_i / \sum_{i=1}^p\lambda_i^2\). We refer to the case when 
\( w=1\) as the PW likelihood.

\subsubsection{Estimating J and H}

To compute the eigenvalues \(\lambda_i\), we estimate the matrices \(H(\tau_0)\) and 
\( J(\tau_0) \) using 
independent internal subsets of the alignment following the empirical estimates of \citet{Cattelan2015}.
Given independent subsequences \((X_i^s), s=1,\dots,S\), the estimates are   
\begin{align*}   
    \hat{H}_{ij} &= \frac{1}{S} \sum_{s=1}^S \sum_{k=1}^{M-1} \sum_{l=k+1}^M \left[ \frac{\partial}{\partial \tau_i} \ell_c( \tau | X^s_{kl}) \right] 
                                                     \left[ \frac{\partial}{\partial \tau_j} \ell_c( \tau | X^s_{kl}) \right], \\
\end{align*}
 \begin{align*}   
     \hat{J}_{ij} &= \frac{1}{S} \sum_{s=1}^S \left[ \frac{\partial}{\partial \tau_i} \ell_c( \tau; X^s) \right]
                                        \left[ \frac{\partial}{\partial \tau_j} \ell_c( \tau; X^s) \right]. \\ 
\end{align*}

As our application is 
long concatenated sequence alignments, we expect that typically enough
independent internal subsets will be available to estimate the  \(J \) and \(H\) matrices. 

Defining the derivatives of the doublet probabilities with respect to the branch lengths
by \(\ppii{d,i}{kl} = \pd{}{\tau_{i}} \pii{d}{kl}\), we have

 \[ 
     \frac{\partial}{\partial \tau_{i}} \ell_c(\vec \tau| X) =
     \underset{(k,l) \in \mathcal{D}_{\tau_i}}{\sum \sum} \sum_{d=1}^D n^{kl}_{d} \frac{\ppii{d,i}{kl}}{\pii{d}{kl}}, 
 \]

 and
 \[ 
     \frac{\partial}{\partial \tau_{i}} \ell_c(\vec \tau| X_{kl}) = \begin{cases}  
         \sum_{d=1}^D n^{kl}_{d} \frac{\ppii{d,i}{kl}}{\pii{d}{kl}} & \text{if } (k,l) \in \mathcal{D}_{\tau_i}\\
        0 & \text{else}.
    \end{cases}
 \]

\subsubsection{Jukes-Cantor adjustment}

For the JC substitution model, all characters are treated identically, so we
only consider whether characters at a site are the same or different. There are
\(D=2\) possible doublet configurations; identical or differing nucleotides.  We
use the notation \textit{11} to indicate identical nucleotides and \textit{10}
for different nucleotides. The doublet probabilities across distance \(\nu\) are
denoted \(p_{10}(\nu) \) and \(p_{11}(\nu)\), and refer to the probabilities of
observing the same nucleotide or not, respectively. We have \(p_{11}(\nu) =
\frac{1}{C} \sum_{i=1}^c \left( \frac{1}{4} + \frac{3}{4} e^{-\frac{4}{3} \nu
r_c} \right) \), \(p_{10}(\nu) = \frac{1}{C} \sum_{i=1}^c \left( \frac{3}{4} -
\frac{3}{4} e^{-\frac{4}{3} \nu r_c} \right). \) 

\[  \ppii{11,i}{kl} = \pd{}{\tau_{i}}  \pii{11}{kl} = - \frac{1}{C} \sum_{c=1}^C \kappa_c e^{-\frac{4}{3} \nu_{kl} \kappa_c} 
\]
\[  \ppii{10,i}{kl} = \pd{}{\tau_{i}}  \pii{10}{kl} =   \frac{1}{C} \sum_{c=1}^C \kappa_c e^{-\frac{4}{3} \nu_{kl} \kappa_c}, 
\]

and the partial derivatives of the PW log likelihood are 
 \[ 
     \frac{\partial}{\partial \tau_{i}} \ell_c( \tau| X) =
     \underset{(k,l) \in \mathcal{D}_{\tau_i}}{\sum \sum} 
        n^{kl}_{11} \frac{\ppii{11,i}{kl}}{\pii{11}{kl}} +
        n^{kl}_{10} \frac{\ppii{10,i}{kl}}{\pii{10}{kl}} .
 \]

The partial derivatives of the pairwise data marginal likelihoods are 
 \[ 
     \frac{\partial}{\partial \tau_{i}} \ell_c( \tau| X_{kl}) = \begin{cases}  
        n^{kl}_{11} \frac{\ppii{11,i}{kl}}{\pii{11}{kl}} +
        n^{kl}_{10} \frac{\ppii{10,i}{kl}}{\pii{10}{kl}} & \text{if } (k,l) \in \mathcal{D}_{\tau_i}\\
        0 & \text{else}.
    \end{cases}
 \]

 We estimate 
 \(\tau_0\) as 
 \[ \tau_0^{kl} = \begin{cases} 
     -\frac{3}{4} \ln ( 1 - \frac{4}{3} p^{kl})                       & \text{for } JC \\ 
     \alpha \left( (1 - \frac{4}{3} p^{kl} ) \right)^{-1/\alpha} - 1  & \text{for } JC+\Gamma 
 \end{cases}
 \]
 where \(p = \frac{n_{10}^{kl}}{n_{10}^{kl} + n_{11}^{kl}}\).

\subsubsection{\(GTR\) adjustment}

The \(GTR\) model allows both the relative rates of substitution between pairs of
nucleotides \(r = (r_{AC}, r_{AG}, r_{AT}, r_{CG}, r_{CT}, r_{GT}) \) and
relative nucleotide frequencies \(\pi = (\pi_A, \pi_C, \pi_G, \pi_T) \) to
differ. Given site mutation rate heterogeneity parameter \(\alpha\), the
nucleotide transition probabilities along distance \(\nu\) are given as the
elements of the probability transition matrix \( P(\nu) = \frac{1}{C}
\sum_{c=1}^C e^{Q \nu \kappa_c}\), where \(Q\) is the typical \(GTR\) rate matrix
defined such that the off-diagonal elements of \(Q\) are the off-diagonal
elements of \( R \Omega\) where \(R\) is the matrix of relative rate parameters and
\(\Omega\) is the diagonal matrix of nucleotide frequencies. The diagonal elements of \(Q\)
are taken such that the row sums of \(Q\) are 0.  The doublet probabilities are
then calculated as \(\Omega P(\nu)\);  their derivatives are not available
analytically, but may be estimated numerically.  We estimate \(\nu_0\) 
for the \(GTR\) and  \(GTR+ \Gamma \) models
using the method of \citet{Waddell1997} for each taxa pair \((k,l)\).

\subsection{Simulations}

We explore the use of the Full, APW1, APW2, and PW likelihoods for 
Bayesian node dating analysis through extensive simulation studies. Then, 
we estimate simulated divergence times under the relaxed clock
model, with six different fossil calibration settings, as
described below. 

The simulations use a 15-taxa subtree of the bird tree topology of
\citet{Wu2024}. Four of the fossils calibrations are used to calibrate the nodes
of the simulation tree, as shown in Figure \ref{fig:sim_fossil}a. The node ages
are generated according to birth-death process prior distribution using the
default speciation and birth rate parameters in MrBayes. Node ages and branch
lengths generated from these prior distributions using an MCMC chain with no
data for 5,000,000 generations with no burn-in, and the median sampled node ages
are used as the "true" node ages. The overall mutation rate (in units of
mutations per millions of years) is assumed to be 0.001, corresponding to
\(10^{-10}\) mutations per site per year. The branch-wise mutation rates are
sampled from independent log-normal distributions, such that each mutation rate 
\(\mu_i\) is drawn independently from a lognormal distribution with mean \(\mu\) 
and variance \( \sigma^2\) \citep{Drummond2006}; the branch lengths in mutation units 
are then \(\tau_i = \mu_i b_i \). We
use \(\sigma = 0.1\). For all simulations, we simulate alignments of length
1,000, 10,000, and 50,000 using the program \texttt{Seq-Gen}
\citep{Rambaut1997}. For each fossil calibration setting, we simulate 100
independent replicate sequence alignments and estimate the node ages with an
MCMC chain of 100,000 generations, discarding the first 50\% of samples as
burn-in and sampling every 200 generations. The adjustment weights for the APW1
and APW2 likelihoods are computed by splitting the alignment into 5 segments of
equal length and taking the elementwise averages of the elements of the \(J\)
and \(H\) matrices from each segment.

To assess the quality of node age estimates we compute the overall root mean
square error \(rMSE = \frac{1}{R D} \sqrt{\sum_{r=1}^R \sum_i (\hat{d}_i^r -
d_i)^2}\), where \(d_i^r\) is the \(i\)th internal node age in the \(r\)th
simulation, and D is the number of internal nodes.  We also compute the
posterior credible interval coverage as \(\frac{1}{R D}  \sum_{r=1}^R \sum_{i=1}^D \mathbb{1}(
\hat{d}_{i,0.025}^r < d_i < \hat{d}_{i,0.975}^r ) \), where
\(\hat{d}_{i,p}^r\) is the \(p\)th percentile of the sampled posterior values
of \(\hat{d}_{i}^r\). To explore how node age estimation performance changes 
when node calibration
prior distributions are changed, for Calibration Settings \(j\neq1\) we compute the
difference in rMSE at a node \(i\) between Calibration Setting \(j\) and Calibration
Setting 1 as rMSE\(_j\) - rMSE\(_1\) where  rMSE\(_j\) = \(\frac{1}{R} \sqrt{\sum_{r=1}^R (\hat{d}_i^r -
d_i)^2}\) with the node age estimates are taken with respect to Calibration Setting 
\(j\).

\subsubsection{Node age estimation }

We consider four simulation settings to examine the quality of node-age
estimates under different prior distribution calibration settings. In all
calibration settings, we use four fossil calibration constraints derived from
the bird analysis. The four calibrations consist of two lower bounds on
internal nodes set with offset exponential distributions, an interval
calibration on the root node set with a uniform distribution, and a calibration
on the second oldest node representing the ancestor of the simulation `bird' clade set by
an offset lognormal distribution. 

\begin{figure}[ht!]
\centering
\begin{minipage}[b]{.9\linewidth}
a)
\end{minipage} \\ 
\begin{minipage}[b]{.9\linewidth}
\centering
\includegraphics[width=0.95\linewidth]{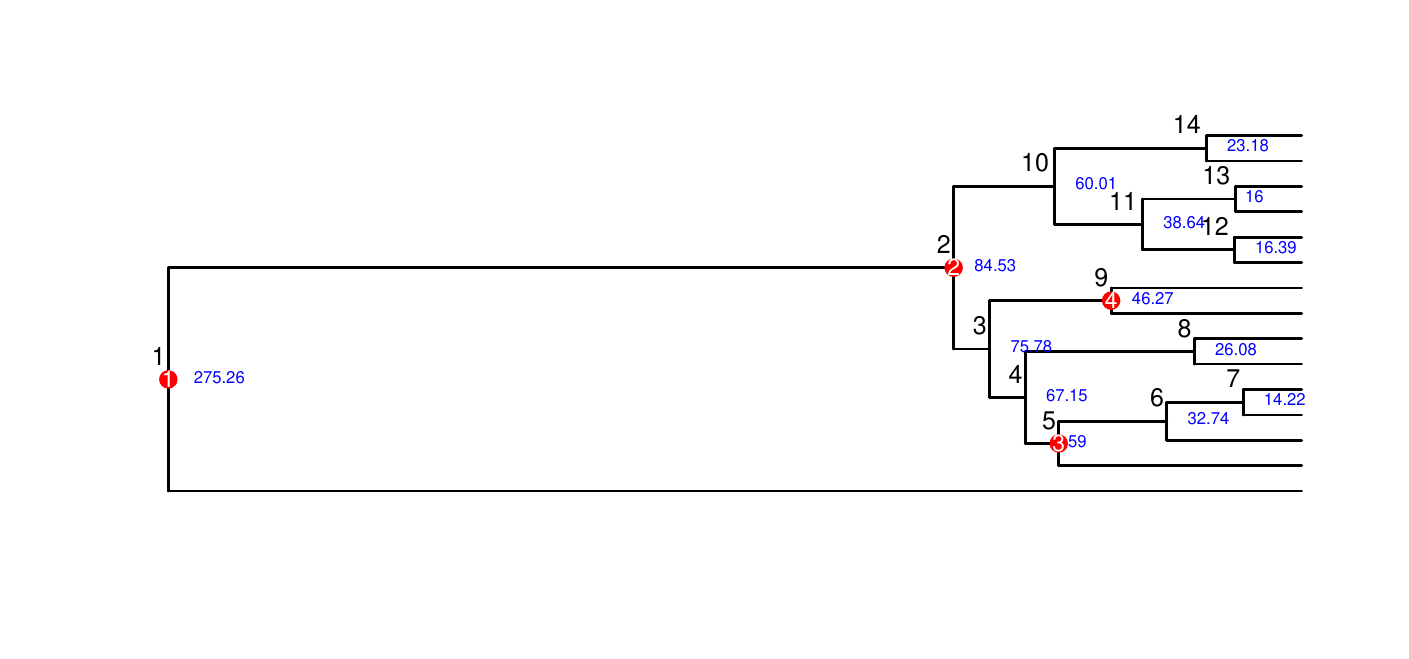}
\end{minipage} \\ 
\begin{minipage}[b]{.9\linewidth}
b)
\end{minipage} \\ 
\begin{minipage}[t]{.9\linewidth}
\centering
\small{
\begin{tabular}{rccccc}
  \hline
\newlinecell{Calibration \\ Setting} & \newlinecell{Setting \\ ID}          & Fossil &  
\newlinecell{Node \\ ID} & \newlinecell{True \\ Node Age} & Prior Calibration             \\  \hline
 \multirow{4}{*}{\newlinecell{Correct  \\ Calibration}} 
 & \multirow{4}{*}{1}   & 1 & 1 &  275.3 & U(255.9, 299.8)       \\ 
 &                      & 2 & 2 &  84.5  & OLN(66.7, 81.3, 5.3)  \\ 
 &                      & 3 & 5 &  59.0  & OE(51.6, 72.8)        \\ 
 &                      & 4 & 9 &  46.27 & OE(28.3, 61.2)        \\  \hline
                       
 \multirow{3}{*}{\newlinecell{Prior \\ Misspecified}}  
 & 2                    & 3 & 5 & 59.0   & OE(61.6, 72.8 )       \\ 
 & 3                    & 2 & 2 & 84.5   & OLN(66.7, 82.6, 9.2)  \\
 & 4                    & 2 & 2 & 84.5   & OLN(66.7, 89.3, 31.8) \\ \hline
                                                           
  \multirow{3}{*}{\newlinecell{Fossil \\ Misplaced}}   
 & 5                    & 3 & 6 & 32.8   & OE(51.6, 72.8)   \\ 
 & 6                    & 3 & 4 & 67.2   & OE(51.6, 72.8)   \\
 & 7                    & 3 & 3 & 75.8   & OE(51.6, 72.8)   \\ \hline
\end{tabular}
}
\caption{
a) 15-taxa tree topology with node ages used for simulations. The Node ID labels are in 
black text, the Fossils are labeled with red circles, and the node ages are in blue text. \\
b) Summary of fossil calibration settings used for node age estimation simulations. 
         In Setting 1, the fossils are placed correctly and the node ages are calibrated 
         using prior distributions that are correct for the true node ages. 
         In Settings 2-4, the prior distributions are changed, and in Settings 
         5-7, Fossil 3 is placed on a different node. For Settings 2-7, the 
         change from the correct fossil setting or prior distribution is indicated. 
         }
\label{fig:sim_fossil}
\end{minipage}
\end{figure}

The quality of fossils used to calibrate the prior distributions of node ages
on the tree is a critical factor in determining the quality of the node age
estimates. Placements of fossil calibrations are often disputed, and there is
considerable uncertainty inherent to the practice of assigning prior
distributions to tree nodes based on the fossil record. We use seven fossil
calibration settings to explore the quality of node age estimates obtained from
the PW, APW1, APW2 and Full likelihoods. In Calibration Setting 1, all fossils
are placed correctly and given prior distributions that are correct for the
true node ages.  In Settings 2, 3, and 4, all four fossils are correctly
placed, but the node ages are calibrated with prior distributions that are
misleading about the true node age (as in the case of Setting 2) or less
informative about the node ages than Setting 1 (as in Settings 2 and 3). In
settings 5,6, and 7, Fossil 3 is placed incorrectly, but the prior distribution
used to calibrate the node age is unchanged. These calibration settings are
discussed in more detail below and summarized in Figure \ref{fig:sim_fossil}b.
We compare the rMSEs obtained in Setting 1 to the rMSEs obtained in Settings
2-7 to explore the impact of changes in the fossil calibration on the node age
estimates obtained by the four likelihoods.

Calibration Setting 1 uses node age calibrations that mimic those used in the bird
divergence analysis, as described in the next section. 
The root node (Node 1) is calibrated with a uniform
distribution with lower bound 255.9 MYA and upper bound 299.8 MYA, denoted
U(255.9, 299.8). The node representing the modern bird ancestor (Node 2) is
calibrated with a lognormal distribution offset to have lower bound 66.7 MYA,
mean 81.3 MYA, and standard deviation of 5.3. This is intended to place 2.5\%
of the distribution's mass above the `soft' upper bound of 94.1, and is
denoted OLN(66.7, 81.3, 5.3). Node 5 is calibrated with an exponential
distribution offset to have lower bound 51.6 and mean 72.8, denoted OE(51.6, 72.8) 
Node 9 is calibrated with an OE(23.3, 61.2) distribution. All calibrations are
appropriate for the true node ages. 

We conduct additional simulation studies comparing the APW, PW, and Full
likelihoods for estimating the branch lengths in evolutionary units for an
unrooted tree topology with no fossil calibrations. For the unrooted tree, we
use the same tree as in the node age simulations in unrooted format.
We also conduct simulations in which the \(GTR+\Gamma\) substitution model
parameters are estimated from the simulated data, rather than fixed to the true
values. These simulations settings are described in the Appendix.

\subsection{Bird divergence analysis}

We apply the pairwise likelihood to a node dating analysis using the bird genome
alignment data of \citet{Wu2024}. The data consist of 100 clock-like protein
coding genes. These genes are subsampled from the 1000 genes with the least
degree of molecular clock violation collected from a whole genome analysis by
\citet{Wu2024}. \citet{Wu2024} collected genomic data from 125 taxa, including
118 modern bird \emph{Neoaves} and an outgroup of 6 ancient birds
\emph{Paleoaves}, and the American alligator. We use 20 fossils to calibrate the
clock model; these fossils are the same as in \citet{Wu2024a}.  For the split
between \textit{Neognathae} and \textit{Paleognathae} (Paleo-Neoave), we
consider three different node age calibrations based on the calibrations used by
\citet{Wu2024}. They use lower and upper bounds in MCMCTree with a soft upper
`bound' that allows the MCMC samples to violate the upper bound with probability
0.025 \citep{Yang_2007}. We use lognormal distributions shifted to have the same
lower bound and with 0.025 of the distribution mass above the soft upper
`bound'. The full fossil calibration details are in the Supplementary Materials.
 In order to compare the results of the node age
estimation using the pairwise likelihood, we assume the same tree topology
estimated in \citet{Wu2024} (Figure \ref{fig:bird_tree})  

For the full likelihood model, parameters are sampled from MCMC chains of
2,000,000 generations, thinned by retaining every 500th sample, and the first
\(50 \%\) of samples are discarded as burn-in. The APW1, and APW2 are run
for 3,000,000 generations with the same settings otherwise. The PW 
likelihoof is run for 4,000,000 generations, as the chains take longer to converge. 
Two independent runs are conducted with 4 Metropolis-coupled chains in each
run. Convergence is monitored by examining trace plots of sampled
log-likelihood values 
(Figure \ref{fig:mcmc_trace}). The chains are run in parallel using OpenMPI
version 5.0.3, with a separate process for each of the 8 MCMC chains.

\subsubsection{Prior distributions}

The prior distributions used for the bird divergence time estimation are as
follows.  The overall mutation rate is given a normal distribution with a mean
of 0.001, standard deviation of 0.01, truncated to be greater than 0. The
qossil ages are
given Offset Exponential distributions with lower bounds
given by the minimum age of the fossil, as described in the Appendix.
The means of the OE distributions 
are set to be at the midpoint of the node lower bound and the soft upper bound 
of the bird ancestral node, the Paleo-Neoave split. The Paleo-Neoave
split is calibrated with an OLN(66.7, 81.3, 5.3) distribution; the
mean and standard deviation are chosen such that the 97.5th percentile matches
the soft upper bound of 94.1 MYA used in \citet{Wu2024} and the mean is halfway
between the lower bound and soft upper bound. The
node ages are given the birth-death process prior distribution as implemented in
MrBayes version 3.2.7, with default priors on speciation and extinction rate
(Exponential with mean 10 and Uniform on [0,1], respectively). We check that the prior 
distributions on the node ages are reasonable by running the MCMC chain without 
data and comparing the resulting 95\% intervals with those obtained 
using MCMCTree and the settings used in \citet{Wu2024}.

\subsubsection{Fixing substitution model parameters}

For the bird divergence time analysis, we fix the substitution model parameters
by using the full likelihood model in a short initial MCMC run. We use the MrBayes 
`propset` command to increase the frequency of relative rate move proposals to 0.1
and \(\alpha\) move proposals to 0.1. The tree
topology is fixed and all model parameters and branch lengths are sampled for
50,000 generations in 2 independent runs, with the first 8,000 discarded as
burn-in. To ensure reasonable values for the substitution models are estimated,
we examine the width of the interval estimates and compare the estimates obtained 
to the parameter estimates obtained in the complete MCMC run using the Full likelihood.

\section{Results}

\subsection{Simulation Results} 

\begin{figure}
    \centering
    \includegraphics[width=0.9\linewidth]{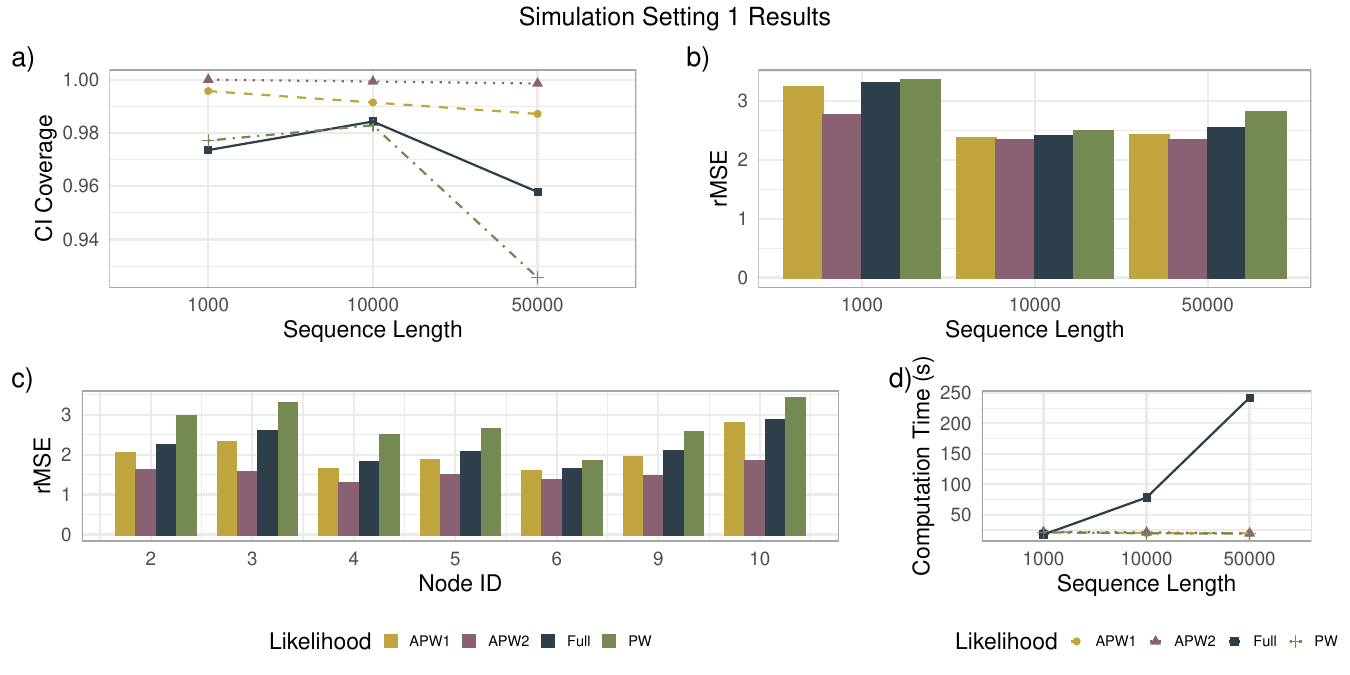}
    \caption{Results of node age simulation Setting 1:  \\
        a) (0.95) Credible Interval coverage rates, averaged across all 
        internal tree nodes; \\
        b) rMSE, averaged across all 
        internal tree nodes; \\        
        c) rMSE for 50,000 site setting at selected nodes;  \\
        d) Computation time in seconds. \\
        The Full likelihood results are represented by the black bars or black
        squares connected by solid lines, the PW likelihood results are
        represented by green bars or green crosses connected by dot-dashed
        lines, the APW1 results are represented by yellow bars or yellow circles
        connected by dashed lines, and the APW2 results are shown by the maroon
        bars or maroon triangles connected by dotted lines. 
}
    \label{fig:cal1_res}
\end{figure}

Under Calibration Setting 1, where calibrated nodes were correctly positioned
and assigned appropriate prior distributions, the four likelihood formulations
exhibited distinct performance patterns in estimating node ages (Figure
\ref{fig:cal1_res}). Assessment of confidence interval (CI) coverage rates
averaged across internal nodes revealed that the APW2 and APW1 likelihoods
achieved the highest overall coverage, both exceeding the nominal 0.95 level
(Figure \ref{fig:cal1_res}a). In contrast, the PW likelihood produced the lowest
coverage, declining to approximately 0.925 under the 50,000-site condition,
although it remained near or above the nominal threshold for shorter alignments.
Square-rooted mean squared error (rMSE) comparisons demonstrated that APW2
consistently yielded the lowest values across all sequence lengths (Figure
\ref{fig:cal1_res}b). The PW likelihood produced the highest rMSE, while APW1
and the Full likelihood displayed comparable performance. Node-specific rMSE
analyses focused on Nodes 2–6 and 9–10, representing the oldest non-root nodes
(\ref{fig:sim_fossil}a) and those most affected by calibration perturbations in
simulation settings 2–7. At each of these nodes, APW1 and APW2 achieved the
lowest rMSE values (Figure \ref{fig:cal1_res}c).

Computational time analyses indicated that all four likelihoods required
approximately 20 seconds to complete MCMC analyses for 1,000-site alignments
(Figure \ref{fig:cal1_res}d). For PW, APW1, and APW2, runtime remained
essentially constant with increasing sequence length. In contrast, the Full
likelihood exhibited a pronounced increase in computational demand, requiring
~75 seconds at 10,000 sites and ~250 seconds at 50,000 sites. Comparable runtime
patterns were observed under Calibration Settings 2–7 and are therefore not
shown.

\begin{figure}[ht!]
    \centering
    \includegraphics[width=0.9\linewidth]{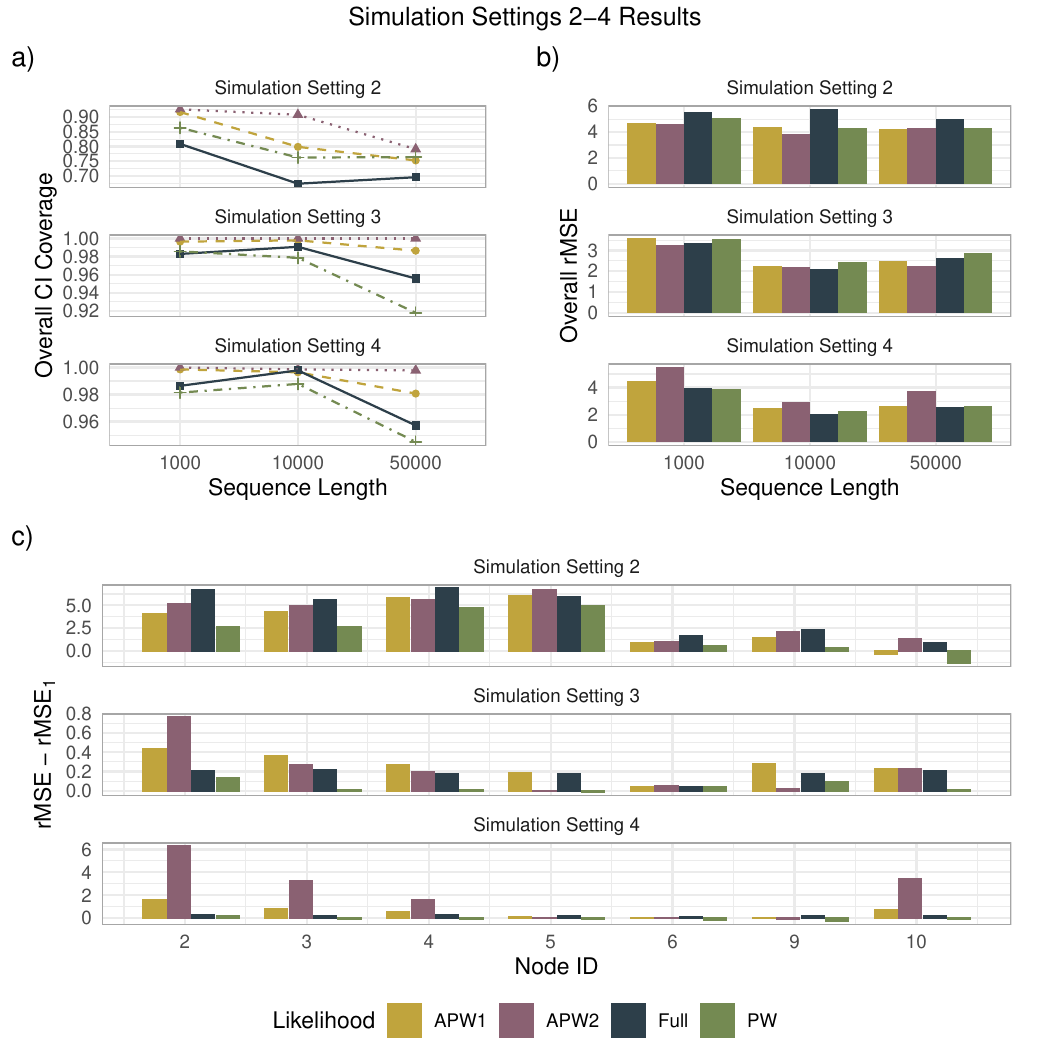}
    \caption{Results of node age simulations where the prior distribution is misspecified,
    Calibration Settings 2-4.   \\
    a) The CI coverage rates, averaged over all internal nodes;  \\
    b) rMSEs,  averaged over all internal nodes; \\
    c) Difference in rMSEs between Settings 2-4 and Setting 1 for selected internal nodes. 
    }
    \label{fig:cal2_res}
\end{figure}

In Calibration Settings 2-4, the fossil placements are unchanged, but are
calibrated with incorrect or less informative priors (Figure
\ref{fig:cal2_res}). The Full likelihood interval coverage rates (averaged over
all internal nodes) are lowest of the four likelihoods in Setting 2 (Figure
\ref{fig:cal2_res}a), in which the prior distribution on the calibration for
Node 5 has a lower bound higher than the true node age. The APW2 likelihood CIs
have the highest coverage rates for all sequence lengths, and the APW1 intervals
have higher coverage rates than the PW likelihood for the 1,000 and 10,000 site
settings, and a similar rate for the 50,000 site setting. The coverage rates for
Calibration Settings 3 and 4 are similar, and are similar to the results of Calibration Setting 1, as
the calibrations are correct for the node ages, but the prior distribution on
the age of Node 2 less informative. Again, the PW likelihood has the lowest
interval coverage rates, and the APW2 and APW1 coverage rates are highest. 

For Calibration Setting 2, the Full likelihood has the highest rMSE overall for
all simulated sequence lengths, and although the overall rMSE for the APW2
compares favorably to the Full likelihood in Setting 3, the APW2 likelihood
overall rMSE is worst in Setting 4. (Figure \ref{fig:cal2_res}b). For Settings 2-4,
we wish to quantify how the change in fossil calibration affects the node age
estimates. We consider the difference between the rMSE obtained in each of
Calibration Settings 2-4 and the rMSE in Calibration Setting 1, denoted rMSE - rMSE\(_1\). The
difference in rMSE represents the change in the quality of the node age estimate
caused by the change in node age calibration prior. Unsurprisingly, the quality
of estimates for Nodes 2-4 is reduced for all likelihoods in Calibration Setting 2 (Figure
\ref{fig:cal2_res}c); the Full likelihood estimates are the most
affected, with the largest change in rMSE at these nodes. The APW1 and APW2
likelihood-based estimates of Node 5 have the largest change in rMSE at that
node. The PW likelihood is the least affected, with the least change in rMSE at
all nodes. For Calibration Settings 3 and 4, the calibration prior on Node 2 is
made less informative. The APW2 based estimates have the largest increases in
rMSE in these settings, with the APW1 likelihoods also showing larger increases
in rMSE than the Full likelihood.  
The interval estimates for all four 
likelihoods achieve low coverage rates for Nodes 2-4 in Calibration Setting 2, but the 
Full likelihood intervals have the lowest coverage rates by far, covering the 
true node ages with a rate of 0 for the 10,000 and 50,000 site settings, the APW1 and 
APW2 likelihood-based intervals cover these true node ages at much higher rates 
(Figure \ref{fig:sim_cal2_all}a). The node-wise changes in rMSE for the 
1,000 and 10,000 site settings are similar to those of the 50,000 site setting 
(Figure \ref{fig:sim_cal2_all}b).

\begin{figure}[ht!]
    \centering
    \includegraphics[width=0.9\linewidth]{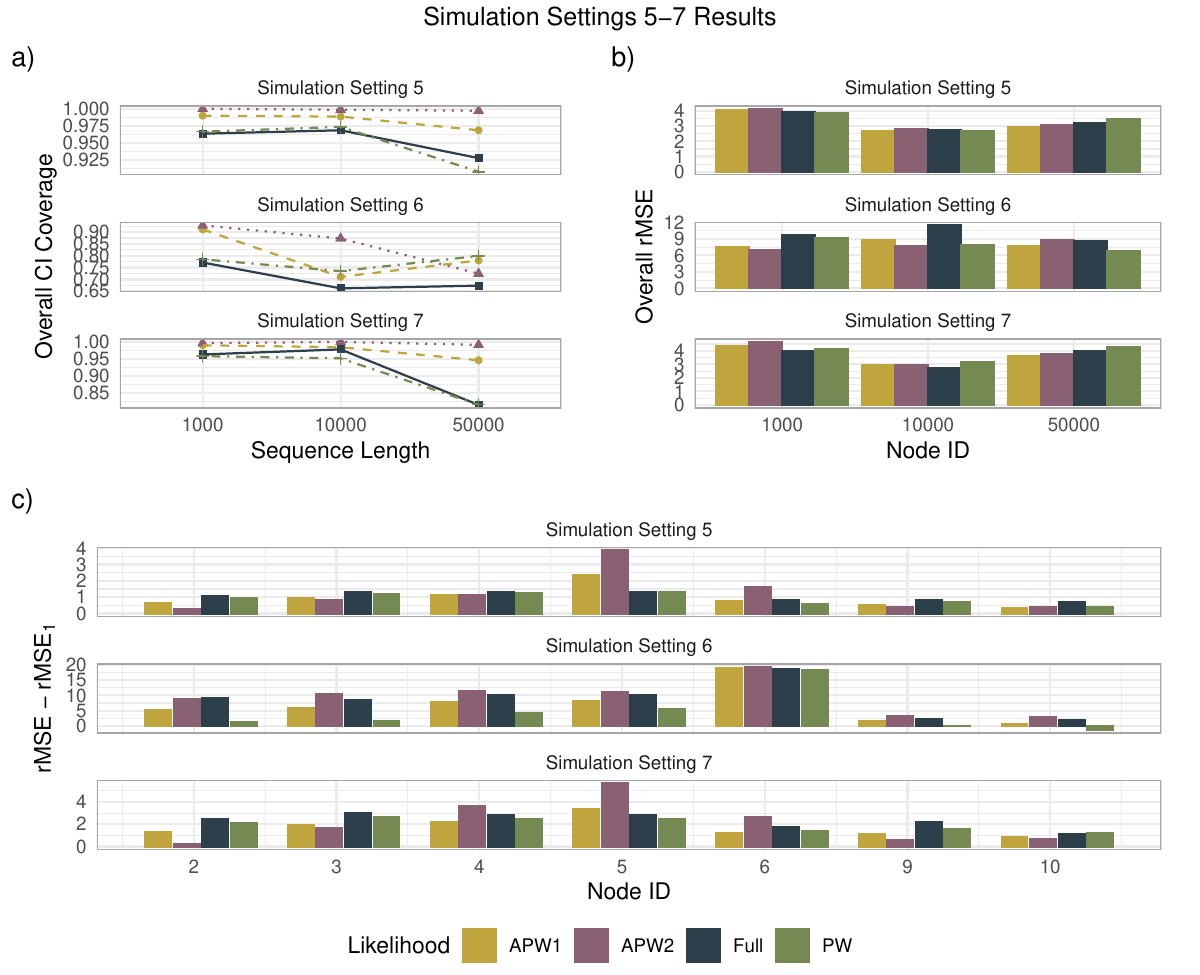}
    \caption{Results of node age simulations in which Fossil 3 is misplaced,
    Calibration Settings 5-7:   \\
    a) The CI coverage rates, averaged over all internal nodes;  \\
    b) rMSE, averaged over all internal nodes; \\
    c) Difference in between rMSE in Setting 5-7 and rMSE in Setting 1 for selected internal nodes.  }
    \label{fig:cal3_res}
\end{figure}

In Calibration Settings 5-7, the
calibration prior distributions are unchanged, but the fossil calibration at
Node 5 is misplaced. The fossil is placed at the direct ancestor and descendant
nodes, 4 and 6 (Settings 5 and 6, respectively), and then at Node 5's
grandparent node, Node 3 (Setting 7). Examining the CI coverage (over all nodes)
shows similar coverage rates across the four likelihoods in 
Calibration Settings 5 and 7, in which the Fossil 3 calibration is
placed above the correct node (Figure \ref{fig:cal3_res}a).
The APW2 and APW1 likelihoods achieve the highest
coverage rates, and the Full and PW likelihood interval coverage rates are
similar. Both Adjusted PW likelihood coverage rates remain above or very near
the credible level of 0.95, but both PW and Full likelihood coverage rates drop
below the credible level when the number of sites is 50,000; in Calibration
Setting 7, both Full and PW likelihood obtain interval coverage rates around
0.825 in the 50,000 site setting. The node age estimates are more affected in
Setting 6, in which the lower bound due to Fossil 3 is placed at Node 4, below
its correct position. Here, the Full likelihood achieves the lowest CI coverage rate
for all sequence lengths. The APW2 likelihood achieves the highest interval coverage rates
in the 1000 and 10,000 site settings, and the PW and APW1 likelihoods achieve the highest
coverage rates in the 50,000 site setting.

In Calibration Settings 5 and 7, the overall rMSEs are similar;
all four likelihoods obtain similar overall rMSEs in the 10,000 site setting, the APW1
and APW2 likelihoods achieve lower rMSEs than the PW and Full likelihoods for 50,000 sites, 
and for 1,000 sites the PW and Full likelihoods achieve slightly lower rMSEs than the two
APW likelihoods. In Setting 6, the Full likelihood overall rMSEs is highest and APW2 is lowest
for 1,000 and 10,000 sites, and the APW1 and PW likelihoods achieve lower rMSEs than the 
Full likelihood for all sequence lengths (Figure \ref{fig:cal3_res}b). 
We again compare the node-wise rMSEs to those obtained in Setting 1; 
the Full likelihood rMSE increases the most at Nodes 2, 3, 9, and 10, 
the APW1 and APW2 likelihood-based rMSEs increase the most at Node 5, and the 
APW2 estimates have the largest increase in rMSE at Node 6 (Figure \ref{fig:cal3_res}c)
In Simulation Setting 6, the APW2 and Full likelihood estimates have similar
increases in rMSE, with 2-6 being the most affected Nodes. The APW1 likelihood
rMSEs increase less than the APW2 and Full likelihoods, and the PW likelihood
node age estimates are the least affected by the fossil misplacement, and obtain
the lowest overall rMSEs in the 50,000 site settings.

The simulations using the unrooted tree suggest the PW, APW1, and APW2 
estimates converge to the true branch lengths as the sequence length increases 
(Table \ref{tab:unr_sim}). The supplementary simulations in which the substitution model 
parameters are estimated from the data indicate there is little effect on the node 
age estimates if the substitution model parameters are known and fixed or estimated
simultaneously (Table \ref{tab:gtr_sim}). 

\subsection{Bird Divergence Analysis}

\begin{figure}[H]
  \centering
  \includegraphics[angle=0,width=1.0\textwidth]{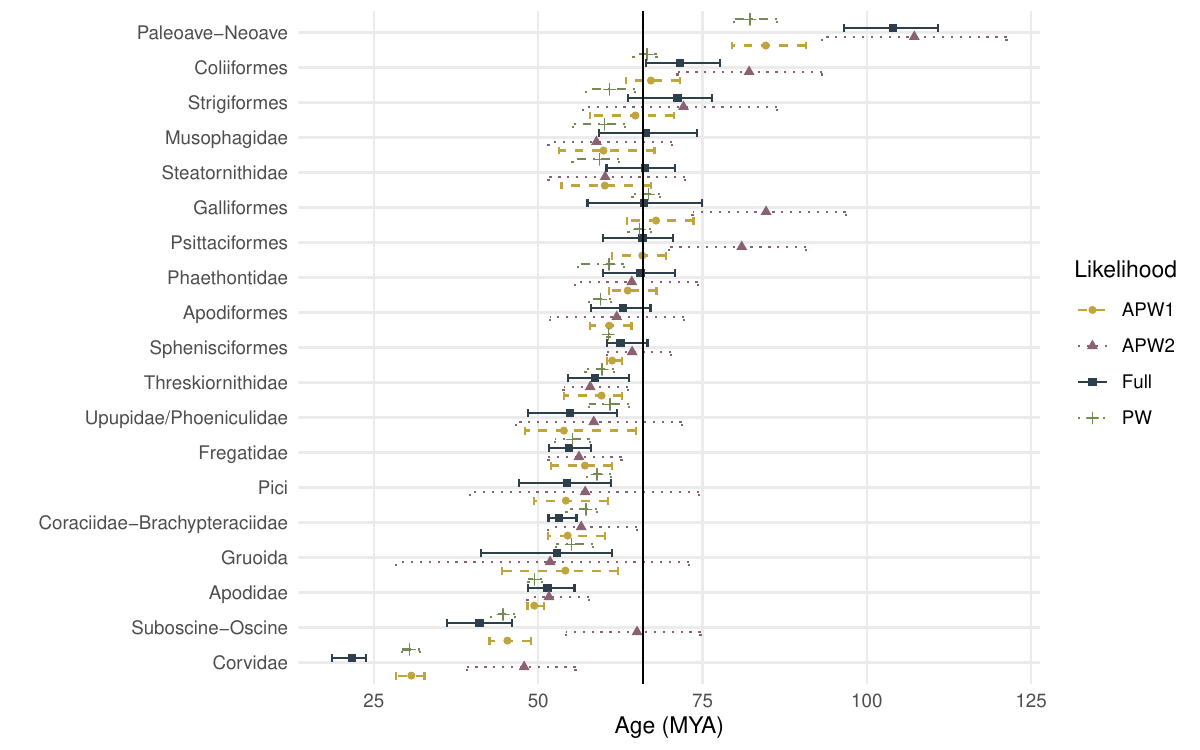}
  \caption{Node age point estimates and credible intervals obtained from the 
   PW (crosses and dotted lines), APW (circles and solid lines), and Full (triangles 
   and dashed lines) likelihoods. The vertical dotted 
  line indicates the kPG boundary, 66 MYA. The nodes on the y-axis are arranged in 
  descending order of the APW point estimate. Note the break in scale on the x-axis,
  to show the estimates of the root node. } 
  \label{fig:node_ages_c1}
\end{figure}

For the nodes where the
fossil calibrations are placed, the interval estimates obtained from the PW
likelihood are the narrowest, and the APW1 intervals tend to be narrower than
the Full likelihood intervals \ref{fig:node_ages_c1}. The APW2 interval
estimates are the widest, consistent with the simulation results. The APW2
likelihood results in some point estimates that are much larger than the other
likelihoods, for the split between Suboscines and Oscines and the root of
Corvidae in the tree in particular; the same is true when considering all estimated node ages
(Figure \ref{fig:all_nodes}). Of interest is whether the radiation of
modern bird species began before or after the kPG boundary; we consider nodes
whose credible interval is entirely above the kPG boundary 66 MYA. The PW
likelihood estimates only 3 such nodes, and the APW1 likelihood estimates 8. The
Full and APW2 likelihood estimate 14 and 17, respectively. The estimated
substitution model parameters fixed for the PW likelihood analyses are shown in
Table \ref{tab:gtr_parm_est}. A comparison with the estimates from the full run of the
node age estimation using the Full likelihood indicates the initial run yields
good estimates of the substitution model parameters.

\begin{figure}[H]
  \centering
  \includegraphics[angle=0,width=1.0\textwidth]{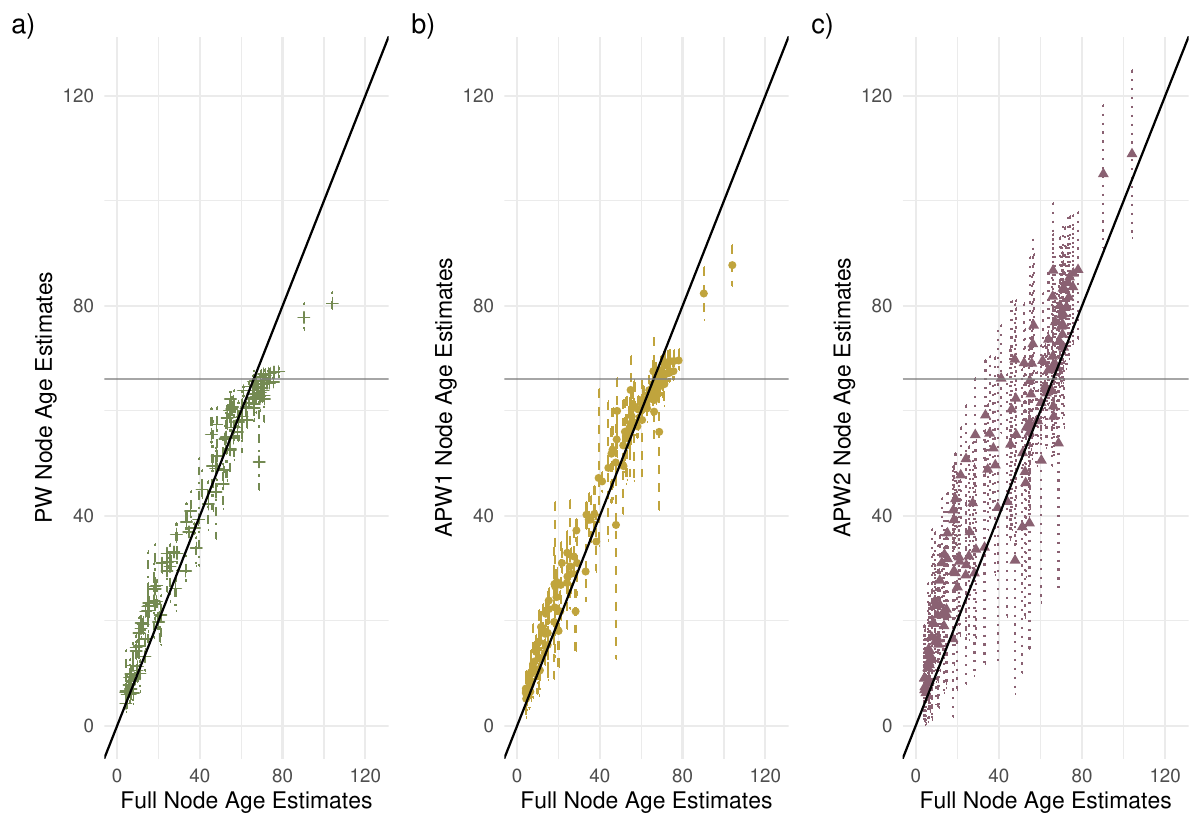}
  \caption{a) Node age estimates obtained from PW and Full likelihood.  \\
  b) Node age estimates obtained from APW1 and Full likelihoods. \\
  c) Node age estimates obtained from APW2 and Full likelihoods.  \\
  The horizontal gray 
  line indicates the kPG boundary, and the black line is the \(x=y\) line. 
  The dot-dashed, dashed, and dotted lines show the PW, APW1, and APW2 credible intervals.}
  \label{fig:plots_pw_full_comp}
\end{figure}

The APW2 likelihood-based estimates have much wider
credible intervals than the PW and APW1 likelihoods, indicating more uncertainty
in the estimated node ages. For the oldest nodes in the tree, the PW and APW1
likelihoods tend to result in younger estimates than the Full likelihood, as
seen in the points near the kPG line falling below the \( x=y \) line 
(Figure \ref{fig:plots_pw_full_comp}). This is
more pronounced in the PW likelihood estimates, whose interval estimates are
also narrower than the APW1 intervals. The APW2 likelihood node age estimates
tend to overestimate the node ages in this region of the tree, but The wider CIs
reflect more uncertainty in the estimates, and tend to cover the Full likelihood
point estimates. For nodes whose estimated age is between 10 MYA and about 50
MYO, all of the PW and APW likelihoods tend to overestimate the node age,
relative to the Full likelihood. However, the wider credible intervals resulting
from the adjustment weighted likelihoods often cover the estimate from the Full
likelihood. Of the 124 internal nodes, the APW2 CI for the node age covers the
point estimate from the Full likelihood for 94, and the number of nodes for
which the credible intervals obtained from the APW2 and Full likelihoods overlap
is 109. The CIs from the APW1 estimates cover the Full likelihood point
estimates and CIs for 81 and 111 nodes, and the CIs obtained from the PW
likelihood overlap the Full likelihood point estimate and intervals for 39 and
91, respectively. 

The MCMC running times for the bird divergence time analysis are given in Table
\ref{tab:comp_times}. The PW and Adjusted PW likelihoods enjoy a considerable
computational advantage over the Full likelihood; with computation times between
119 and 167 minutes, the PW models are on average 27 times faster than the Full
likelihood models, which take between 3412 and 4043 minutes to complete. 
The initial run used to estimate the substitution model parameters
uses 246 minutes; even accounting for this, and the longer MCMC chains, 
the total running times for the PW and APW 
likelihoods are substantially faster than the Full likelihood.  

\begin{table}[ht!]
    \centering
    \begin{tabular}{cccc} 
    & \multicolumn{3}{c}{MCMC Run Time (Minutes)} \\
    Likelihood &  Main Run  & Initial Run & Total  \\ \hline
       PW      &     278.0  & 240.2 & 518.2 \\
       APW1    &     163.4  & 240.2 & 403.6 \\
       APW2    &     245.8  & 240.2 & 486.0 \\      
       Full    &     6204.0 & NA & 6204.0  \\ \hline
    \end{tabular}
    \caption{Computation times for bird divergence time analysis. The initial run is used to estimate the 
    substitution model parameters, which are fixed in the PW and APW runs. For the PW likelihood, the main 
    run takes longer to converge, using 4,000,000 generations}
    \label{tab:comp_times}
\end{table}

In the bird alignment data example, our results using the APW2 likelihood are
similar to the results of \citet{Wu2024}, which is not unexpected since we use
the same fossil calibrations and attempt to calibrate the priors to mimic the
use of MCMCTree with soft constraints. It is known that different software
packages construct the node age prior differently given the same fossil
calibrations, which affects node dating results \citep{BarbaMontoya2017}.
However, the results of our analysis with the APW2 and Full likelihoods arrive
at similar conclusions that the radiation of modern birds began before the kPG
boundary. Our node age estimates from both the APW2 and Full likelihood are
somewhat younger than theirs, however. The results of the APW1 and APW2
estimates differ somewhat; although in the simulations the APW1 behaves
similarly to the Full likelihood, in the bird data analysis, the APW2 has better
agreement with the Full likelihood, and the APW1 and PW likelihood based results
are more similar, and favor more recent divergences. Although the simulation
results indicate both adjustment weights work well if the prior and data agree,
the APW2 leads to wider credible intervals and greater uncertainty in the
posterior estimates, particularly for younger nodes. However, the simulations
indicate that the credible intervals (of all four likelihoods) fail to
adequately reflect the uncertainty in the estimated fossil ages in many cases.
Ultimately, accuracy of the fossil calibrations is certainly the most critical
element in obtaining accurate estimates from the posterior distribution of node
ages, despite the choice of likelihood used. 

\section{Discussion}

Accurate estimation of node ages from phylogenomic data remains a considerable
challenge. The pairwise likelihood (PW) and its adjusted variants (APW1 and
APW2) offer substantial computational advantages over the full-likelihood
approach because their complexity is independent of sequence length. For
example, in the avian dataset, the primary MCMC analysis using PW and APW was
approximately 25 times faster than the full-likelihood method, with similar
gains observed in simulation studies. This advantage becomes increasingly
pronounced for longer sequences, making APW particularly advantageous when
analyzing concatenated alignments. Beyond computational efficiency, our
simulations reveal a fundamental trade-off between estimation precision and
methodological robustness. The full-likelihood approach produced the most
accurate point estimates of node ages; however, this precision was accompanied
by pronounced sensitivity to prior misspecification, leading to substantial
performance degradation under scenarios involving fossil calibration errors. In
contrast, PW demonstrated strong robustness to such inaccuracies, albeit at the
cost of reduced precision. The adjusted pairwise likelihoods (APW1 and APW2)
effectively reconciled these competing properties. By retaining the accuracy of
the full-likelihood method while incorporating the robustness of PW, the
adjusted approaches achieved a balance between precision and
stability. Consequently, APW1 and APW2 consistently outperformed both original
methods, providing the most reliable interval coverage of true node ages across
all simulation conditions.

Genomic-scale phylogenetic datasets are often
comprised of data from many genes, and it is perhaps not realistic to assume a
single substitution process is shared across all possible model partitions. On
the other hand, models with many partitions require many parameters, which can
render model estimation and MCMC convergence problematic. Under-partitioned
models can perform capably for node age estimation if the prior distributions
and the clock model are not badly violated \citep{Angelis_2017}. In the case of
genomic data, many loci may have significant congruence, and concatenating loci
into longer sequences can provide a simpler modeling framework
\citep{Leigh_2008}. The use of the pairwise likelihood then seems reasonable in
settings when the data can be grouped into a few similar, long partitions. Use
of a variable site-rate model like the GTR\(+\Gamma\) can compensate somewhat
for differing substitution processes at different sites and realistic, highly
parameterized models that fit different substitution processes often have
difficulty converging.    

The most difficult aspect of the pairwise likelihood is computing the adjustment
weight. We focus on two of the most simple proposed weights, and we estimate
them only via estimates of the matrices \(J\) and \(H\) obtained by taking
subsets of the existing alignment. In the case of the simulated data, these data
subsets are clearly appropriate, as the model from which the data are simulated
is known; in the case of real data however, the assumption of independent and
identical data partitions may not be met. In our observation, the APW2
adjustment weight is a smaller constant than the APW1 weight, which has the
effect of \`flattening' the APW2 posterior distribution, and can result in
too-wide credible intervals, as seen in the simulation CI coverage rates above
the credible level. The adjustment weights are possibly more problematic to
compute in this case than in other applications of the composite likelihood, as
the weight estimate is of much higher dimension than in typical applications
\citep{Varin2011, Ribatet2012, Cattelan2013, Cattelan2015}. Using simulated
sequences to approximate the weight as in \citet{Cattelan2013} may provide more
robust estimates of the adjustment weights, and using lower dimensional
approximations to the \(H\) and \(J\) matrix or approaches based on
down-weighting less informative taxa pairs as in \citet{Holder2011} may be an
interesting approach. To simplify the adjustment weight computation, we assume fixed 
parameters for the \(GTR\) substitution model.  

Our simulations show the pairwise likelihood can be used for fast and accurate
node age estimation, under different substitution model settings and settings
where fossil calibrations provide inaccurate prior information. However, our
study focuses on a single relaxed clock model and a single 15-taxa tree
topology. More extensive simulation using more varied tree topologies, clock
model or substitution model parameter settings could be undertaken to examine
the behavior of the PW and APW likelihoods in realistic scenarios. Further work
could use examine the performance of node age estimation when other aspects of
the model like the clock model form are misspecified, or make use of a Bayes
factor \citet{Kass1995} based on the APW likelihood to apply the APW to model selection.   

Node dating is a challenging problem in phylogenetic analysis. Increasing
availability of genomic data poses a challenge, as likelihood based statistical
methods can be computationally prohibitive for large datasets. We employ the
pairwise composite likelihood in a Bayesian model to address this problem for
long sequences; since computation of the pairwise composite likelihood doesn't
scale with sequence length, it can be applied to large sequence alignments when
the full likelihood would be infeasible. We propose adjusting the pairwise
likelihood with an adjustment weight derived from the asymptotic form of the
composite LRT; our simulation results indicate this adjustment compensates for
the simplifying assumptions of the composite likelihood, and results in
posterior interval estimates that better reflect the uncertainty in the
estimated node ages. Additionally, our simulation results suggest the 
pairwise likelihood is robust to fossil miscalibrations when estimating 
long-ago divergence times, an intriguing feature as there is considerable  
uncertainty and disagreement in the placement of fossils.

\section{Software and Data Availability}

The pairwise likelihood is implemented as a source Bayesian phylogenetic
software \texttt{MrBayes} version 3.2.7. The implementation is designed not to
interfere with the other features of MrBayes, so the pairwise likelihood and
adjustment weights are available to use with a wide variety of existing
phylogenetic models. As of now, the pairwise likelihood is available only for
DNA type data, and the GTR and JC substitution models. The PW and APW
likelihoods are implemented to work with existing MrBayes functionality,
including MPI parallelization, model partitioning, and stepping-stone sampling
\citep{Xie2010}. The source code is available for compilation from
\texttt{https://www.github.com/gmellison/PwBayes}, which also includes scripts
used for the simulations. No new data were used for this study. Bird genomic
were previously made available at https://doi.org/10.6084/m9.figshare.21499230.v1
\citep{https://doi.org/10.6084/m9.figshare.21499230.v1} and https://doi.org/10.26036/cnp0002307
\citep{https://doi.org/10.26036/cnp0002307}.

\section{Acknowledgments} 

We would like to thank the authors of \citep{Wu2024} for helpful discussion and
for providing their tree topology and sequence data. This study was supported in
part by resources and technical expertise from the Georgia Advanced Computing
Resource Center.

\bibliographystyle{plainnat}
\bibliography{references}



\newpage
\appendix 
\section{Appendix: Supplementary Materials and Methods} 
\subsection{Branch length estimation simulation}

In the first simulation we estimate the branch lengths in mutation units,
assuming an unrooted tree topology. Alignments are simulated under the \(JC\) and
\(GTR\) substitution models, both with independent and identically distributed (iid)
mutation rates at each site, and with variable rates distributed according to
the Discrete Gamma distribution (\(JC+\Gamma\), \(GTR+\Gamma\)) with 4 rate
categories and rate parameter \(\alpha = 0.5\) \citep{Yang1996a}. For each
simulation setting, we generate 500 independent replicates. 

We estimate the branch lengths after fixing the tree topology and substitution
rate parameters at the true values. For the \(GTR+\Gamma \), we also estimate
the branch lengths after conducting a short MCMC run to estimate the relative
rate and \(\alpha\) parameters, and fixing the values to the posterior mean for
the longer MCMC run used to estimate the branch lengths.

For the simulation settings described above, we assume the true substitution
model parameters (including the site-rate heterogeneity parameter \(\alpha\))
are known. In practice, these parameters may be estimated and fixed prior to
estimating the branch lengths or node ages. We simulate applying this procedure
to both the branch length and node age estimation simulations using the GTR
substitution model with \(\alpha=0.5\). We conduct an initial MCMC run of 5000
generations with a burn-in of 1000 generations, in which only the substitution
model parameters are sampled. The posterior mean is then fixed for the main MCMC
run in which the branch lengths or node ages are sampled.

\subsubsection{\(GTR + \Gamma\) parameter estimation}

For use of the PW and APW likelihoods, we assume the site rate heterogeneity
parameter \(\alpha\) is fixed. Additionally, the derivation of the adjustment
weights for the APW1 and APW2 likelihoods assumes the substitution model
parameters are fixed. We conduct additional branch length and node 
age estimation simulations using the \(GTR+\Gamma\) 
substitution model in which we estimate the substitution model parameters from the 
data. For the PW, APW1, and APW2 likelihoods, we conduct a short initial 
MCMC run using the Full likelihood, from which estimates of the GTR relative 
rate parameters and \(\alpha\) are obtained, and used to estimate the adjustment weight 
for the APW1 and APW2 cases. The estimate of \(\alpha\) is then fixed, but the 
substitution model rate parameters are sampled during a main MCMC run to estimate
the branch lengths or node ages. We compare these to estimates obtained by the Full
likelihood in which all the \(GTR + \Gamma\) parameters are sampled to ensure the 
estimates from the initial run are reasonable.

\subsection{Fossil Calibrations} \label{appendix_fossils}

\begin{enumerate} 

\item Split between lizard and chicken. Uniform(255.9, 299.8) prior distribution
\citep{Benton_2009}. 

\item Split between \textit{Paleognathae} and
\textit{Neognathae}. OLN(66.7, 81.3, 5.3), which uses \textit{Asteriornis maastrichtensis} as the
lower bound and \textit{Icthyornis Dispar} as the `soft' upper bound.
\citep{Field_2018, Field_2020}. 

\item Stem \textit{Corvidae}. OE(7.2, 50.6) based on fossil \textit{Corvus larteri}  \citep{MilneEdwards}.

\item Split between suboscines and oscines. OE(13.6, 53.8) based on fossil \textit{Miocitta galbreathi} \citep{Brodkorb1972}.

\item Stem \textit{Psittaciformes}. OE(53.5, 73.8) based on fossil \textit{Pulchrapollia gracilis} \citep{Dyke_2000}.

\item Stem \textit{Strigiformes}. OE(56.8, 75.4) based on fossils \textit{Berruornis orbisantiqui}, \textit{Ogygoptynx wetmorei} \citep{Mayr_2002, Mourer‐Chauvire1994, Rich1976}.

\item Stem \textit{Gruoidea}. OE(28.3, 61.2) based on fossil \textit{Parvigrus pohli} \citep{Mayr2005}.

\item Stem \textit{Apodiformes}. OE(51.58, 72.8) based on fossil \textit{Eocypselus rowei} \citep{Mayr2003, Ksepka2013}.

\item Stem \textit{Apodidae}. OE(48.4, 71.2) based on fossil \textit{Scaniacypselus wardi} \citep{Ksepka2013, Mlikovsky2002}.

\item Stem \textit{Steatornithidae}. OE(51.58, 72.8) based on fossil \textit{Prefica nivea} \citep{Olson1987, Nesbitt_2011, Mayr2005, Harrison1984}.

\item Stem \textit{Sphenisciformes}. OE(60.5, 77.3) based on fossil \textit{Waimanu manneringi} \citep{Slack_2006}.

\item Stem \textit{Fregitidae}. OE(51.58, 72.8) based on fossil \textit{Limnofregata azygosternon} \citep{Olson1977, Smith2010}.

\item Stem \textit{Threskiornithidae}. OE(53.9, 74) based on fossil \textit{Rhynchaeites} \citep{Mayr2011}.

\item Stem \textit{Phaethontidae}. OE(55.6, 74.8) based on fossil \textit{Lithoptila abdouensis} \citep{Bourdon2005}.

\item Stem \textit{Musophagidae}. OE(51.58, 72.8) based on fossil \textit{Foro panarium} \citep{Field_2014, Olson1992}.

\item Stem \textit{Coliiformes}. OE(56.22, 75.2) based on fossil \textit{Sandcoleus copiosus} \citep{Houde1992, Ksepka_2009}.

\item Stem \textit{Coraciidae} and \textit{Brachypteraciidae}. OE(51.57, 72.8) based on fossil \textit{Primobucco mcgrewi} \citep{Ksepka_2010, Brodkorb1972, Clarke2009}.

\item Stem \textit{Pici}.  OE(28.3, 61.2) based on fossil \textit{Rupelramphastoides knopfi}  \citep{Mayr2005, Mayr2006a}.

\item Stem \textit{Upupidae} and \textit{Phoeniculidae}.  OE(46.6, 70.3) based on fossil \textit{Messelirrisor grandis} \citep{Mayr2000, Mayr2006}.

\item Stem \textit{Galliforms}. OE(51.58, 72.8) based on fossil \textit{Gallinuloides wyomingensis} \citep{Ksepka_2009}.

\end{enumerate}

\newpage
\subsection{Bird Tree Topology} \label{appendix:tree}
\begin{figure}[ht!]
  \centering
  \includegraphics[angle=0,width=0.9\textwidth]{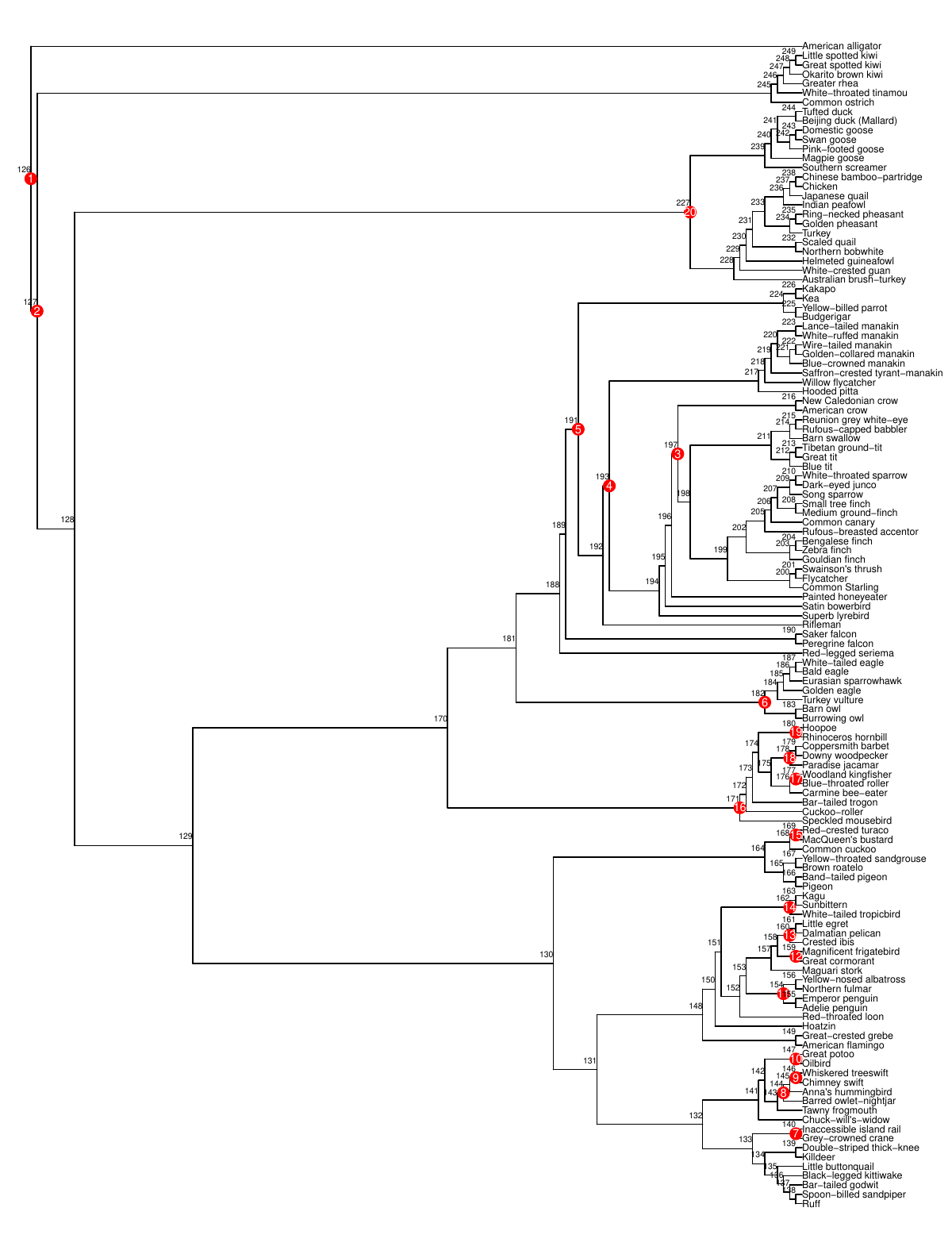}
  \caption{Tree topology estimated from 100 concatenated genes. Nodes calibrated by 
  fossil information are indicated by red points.} 
  \label{fig:bird_tree}
\end{figure}

\section{Appendix: Supplementary Results}

\begin{table}[ht]
\centering
\begin{tabular}{llrrrrrr}
& & \multicolumn{3}{c}{rMSE} & \multicolumn{3}{c}{CI Coverage} \\ \cmidrule(lr){3-5} \cmidrule(lr){6-8} 
\multirow{2}{*}{Likelihood} & \multirow{2}{*}{\newlinecell{Sub \\ Mod}} & \multicolumn{3}{c}{Sites} & \multicolumn{3}{c}{Sites} \\ 
& & 1000 & 10000 & 100000 & 1000 & 10000 & 100000 \\ \cmidrule(lr){3-5} \cmidrule(lr){6-8}
  Full      & \(JC\)             & 0.007450 & 0.002343 & 0.001090 & 0.932 & 0.937 & 0.931  \\
  PW        & \(JC\)             & 0.007414 & 0.002391 & 0.001122 & 0.865 & 0.868 & 0.860  \\
  APW1      & \(JC\)             & 0.007378 & 0.002388 & 0.001125 & 0.976 & 0.979 & 0.976  \\
  APW2      & \(JC\)             & 0.007913 & 0.002491 & 0.001137 & 0.995 & 0.997 & 0.996  \\ 
  Full      & \(JC+\Gamma\)  & 0.010303 & 0.003048 & 0.001420 & 0.935 & 0.938 & 0.936  \\
  PW        & \(JC+\Gamma\)  & 0.010644 & 0.003255 & 0.001521 & 0.873 & 0.879 & 0.876  \\
  APW1      & \(JC+\Gamma\)  & 0.010524 & 0.003274 & 0.001515 & 0.976 & 0.986 & 0.986  \\
  APW2      & \(JC+\Gamma\)  & 0.012393 & 0.003725 & 0.001750 & 0.992 & 0.999 & 0.999  \\ 
  Full      & \(GTR\)            & 0.008081 & 0.002384 & 0.001076 & 0.935 & 0.935 & 0.930  \\
  PW        & \(GTR\)            & 0.007877 & 0.002432 & 0.001116 & 0.866 & 0.868 & 0.863  \\
  APW1      & \(GTR\)            & 0.007864 & 0.002436 & 0.001120 & 0.972 & 0.980 & 0.978  \\
  APW2      & \(GTR\)            & 0.008503 & 0.002577 & 0.001130 & 0.993 & 0.997 & 0.997  \\
  Full      & \(GTR+\Gamma\) & 0.010928 & 0.003244 & 0.001513 & 0.931 & 0.936 & 0.933  \\
  PW        & \(GTR+\Gamma\) & 0.010779 & 0.003367 & 0.001600 & 0.868 & 0.876 & 0.870  \\
  APW1      & \(GTR+\Gamma\) & 0.010787 & 0.003401 & 0.001600 & 0.973 & 0.984 & 0.985  \\
  APW2      & \(GTR+\Gamma\) & 0.012566 & 0.003998 & 0.001856 & 0.991 & 0.998 & 0.999  \\  \hline     
\end{tabular}   
\caption{Simulation node age estimation results. The bottom three rows
show the results of the node age estimation when the \(GTR\) rate parameters are
estimated simultaneously (Full likelihood) or estimated with a short run using
the full likelihood and fixed for subsequent analysis (Pairwise, Adjusted
Pairwise)} \label{tab:unr_sim} \end{table}

\begin{table}[ht!]
\centering
\begin{tabular}{llcrlrrrrrr}
& & & \multicolumn{3}{c}{rMSE } & \multicolumn{3}{c}{CI Coverage }       \\  \cmidrule(lr){4-6} \cmidrule(lr){7-9} 
& & & \multicolumn{3}{c}{Sites } & \multicolumn{3}{c}{Sites }            \\
 &  Likelihood  &  & 1000 & 10000 & 50000 & 1000 & 10000 & 50000  \\ \cmidrule(lr){2-2} \cmidrule(lr){4-6} \cmidrule(lr){7-9}  
\multirow{4}{*}{\newlinecell{Branch  \\ Lengths}} &  
    Full      & & 0.01106 & 0.00356 & 0.00151 & 0.94 & 0.94 & 0.93 \\   
 &  PW        & & 0.01109 & 0.00375 & 0.00159 & 0.89 & 0.89 & 0.87 \\  
 &  PWA1      & & 0.01110 & 0.00370 & 0.00162 & 0.98 & 0.99 & 0.98 \\  
 &  PWA2      & & 0.02885 & 0.00817 & 0.00311 & 0.98 & 1.00 & 1.00 \\   
 & & & \multicolumn{3}{c}{Sites } & \multicolumn{3}{c}{Sites }            \\
 &  Likelihood  & \(\nu\) Fixed & 1000 & 10000 & 50000 & 1000 & 10000 & 50000  \\ \cmidrule(lr){2-2} \cmidrule(lr){3-3} \cmidrule(lr){4-6} \cmidrule(lr){7-9}  
 \multirow{8}{*}{\newlinecell{Node \\ Ages}}   
    &  Full      & No   & 3.51 & 2.50 & 2.40  & 0.97 & 0.99 & 0.97   \\  
    &  Full      & Yes  & 3.45 & 2.44 & 2.31  & 0.98 & 0.98 & 0.98   \\  
    &  PW        & No   & 3.60 & 2.91 & 3.06  & 0.97 & 0.95 & 0.86   \\  
    &  PW        & Yes  & 3.56 & 2.73 & 3.22  & 0.98 & 0.98 & 0.89   \\  
    &  PWA1      & No   & 3.37 & 2.71 & 3.02  & 0.99 & 0.98 & 0.94   \\ 
    &  PWA1      & Yes  & 3.40 & 2.66 & 2.72  & 0.99 & 0.99 & 0.98   \\  
    &  PWA2      & No   & 2.23 & 1.80 & 1.91  & 1.00 & 1.00 & 1.00   \\  
    &  PWA2      & Yes  & 2.33 & 1.78 & 1.92  & 1.00 & 1.00 & 1.00   \\  \hline
\end{tabular}
   \caption{Results for simulations where \(GTR+\Gamma\) parameters (\(\nu\)) are 
   estimated from the data. The top region of the table shows the results of branch length
   estimation in mutation units. The bottom region shows the results of the node age 
   estimation. For the node age estimation setting, the `\(\nu\) Fixed' column indicates
   whether the \(GTR+\Gamma\) parameters are fixed after the initial MCMC run or 
   sampled during the main MCMC run. }
   \label{tab:gtr_sim}
\end{table}

\begin{figure}
    \centering
    \includegraphics[width=0.9\linewidth]{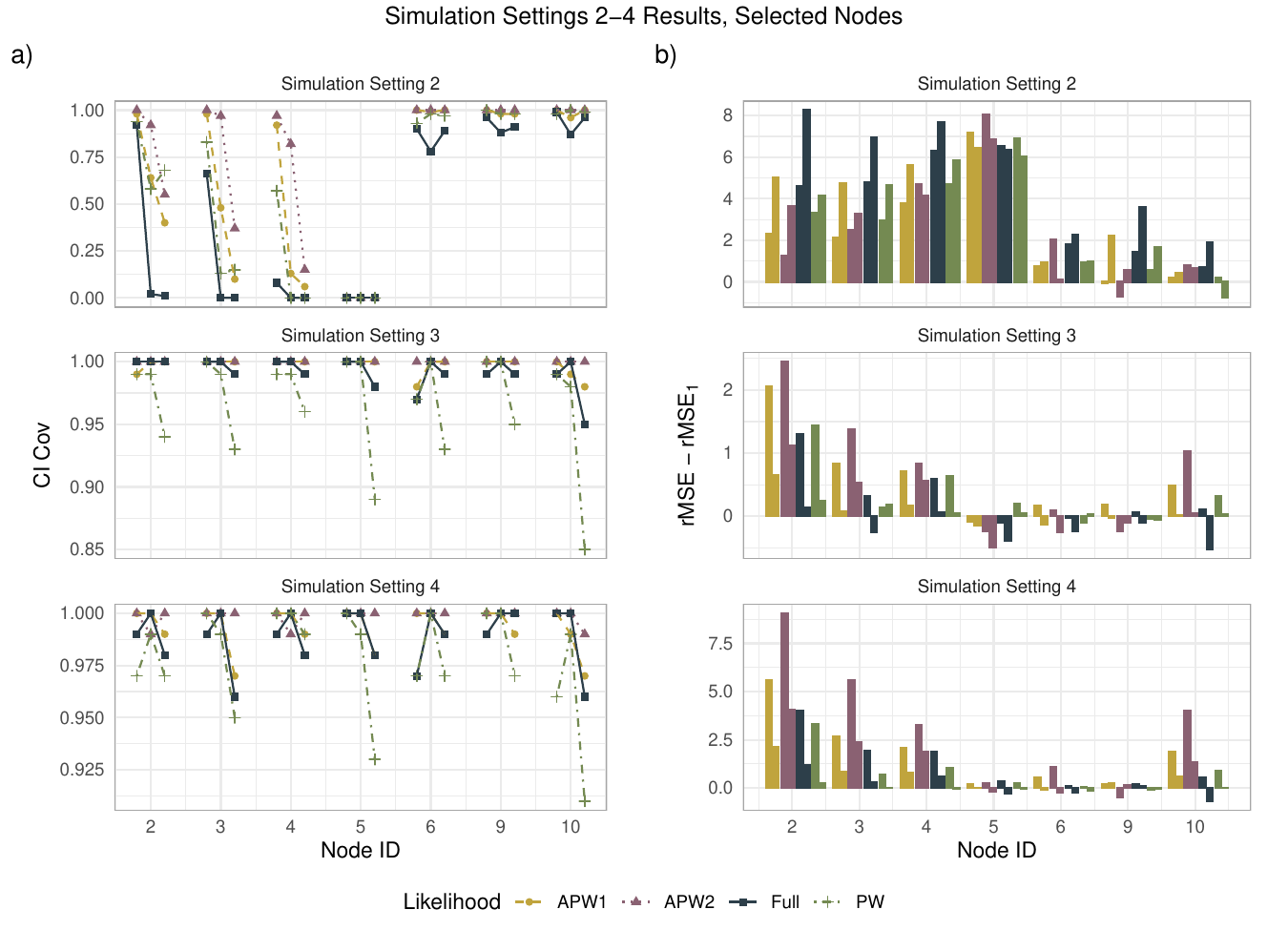}
    \caption{
    Node-wise CI results for Simulations 2-4, selected nodes.  \\
    a) Node-wise CI Coverage for Calibration Settings 2-4.  \\ 
    b) Node-wise difference in rMSE between Calibration Settings 2-4 and Setting 1.  \\
       For each node, there are 2 bars for each likelihood that represent (from left to right) 
       the 1000 site and 10,000 settings. 
    }
    \label{fig:sim_cal2_all}
\end{figure}

\begin{figure}
    \centering
    \includegraphics[width=0.95\linewidth]{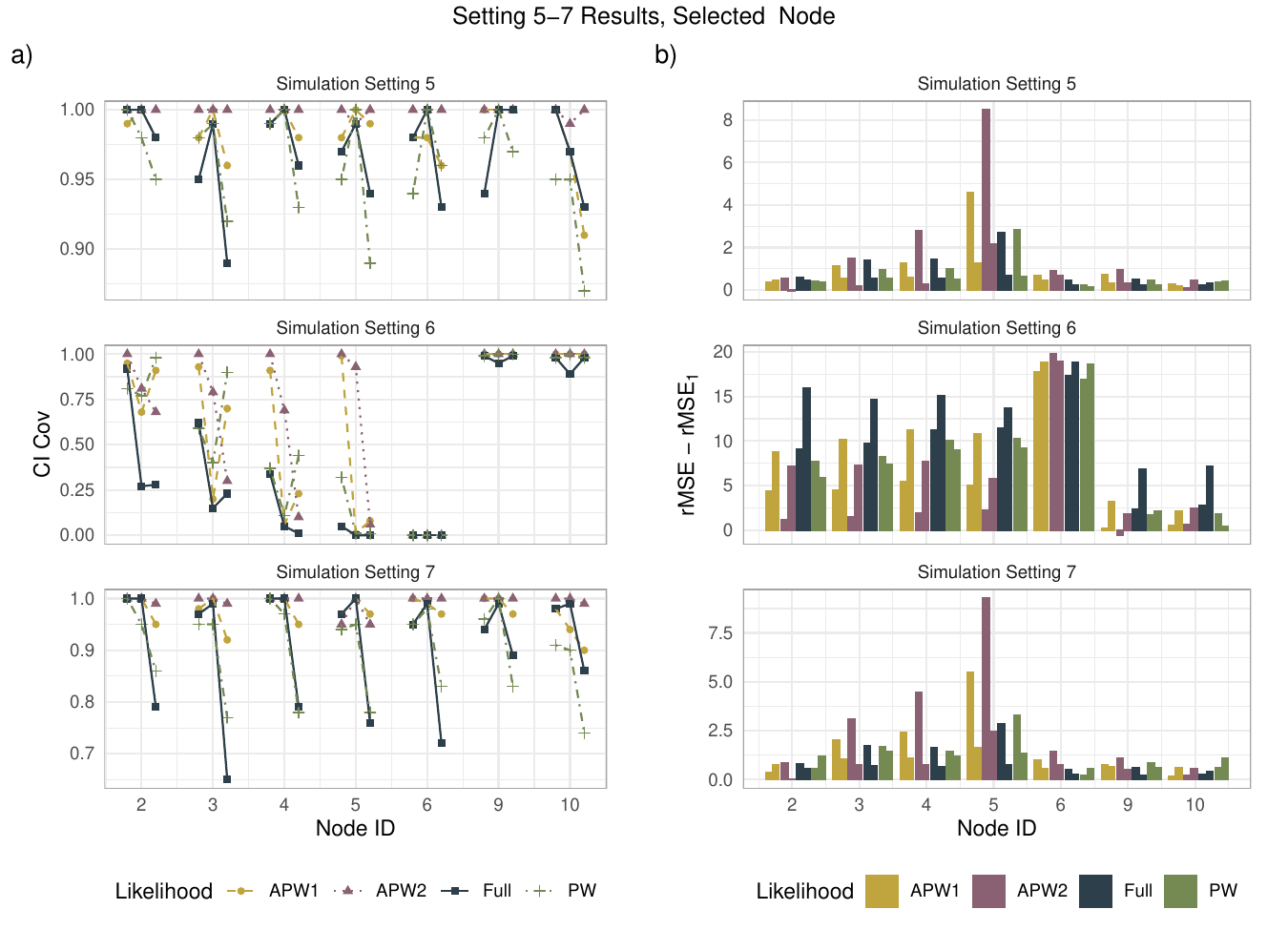}
    \caption{
    Node-wise CI results for Simulations 5-7, selected nodes.  \\ 
    a) Node-wise CI Coverage for Calibration Settings 5-7.  \\
    b) Node-wise difference in rMSE between Calibration Settings 5-7 and Setting 1.  \\
       For each node, there are 2 bars for each likelihood that represent (from left to right) 
       the 1000 site and 10,000 settings. }
    \label{fig:sim_cal3_all}
\end{figure}

\begin{table}[ht!]
    \centering
    \begin{tabular}{ccc}
       Parameter & Full Likelihood Run  & Initial Run \\
       \(r_{AC}\)  & 0.0812 & 0.0822 \\
       \(r_{AG}\)  & 0.3670 & 0.3278 \\
       \(r_{AT}\)  & 0.0437 & 0.0445 \\
       \(r_{CG}\)  & 0.1122 & 0.1116 \\
       \(r_{CT}\)  & 0.3706 & 0.3697 \\
       \(r_{GT}\)  & 0.0652 & 0.0653 \\
       \(\alpha\)  & 0.4495 & 0.4525 \\ 
       \end{tabular}
    \caption{\(GTR+\Gamma\) parameters estimated from Full likelihood run and fixed for PW 
    likelihoods using a short initial run. }
    \label{tab:gtr_parm_est}
\end{table}

\begin{figure}[ht!]
  \centering
  \includegraphics[angle=0,width=0.9\textwidth]{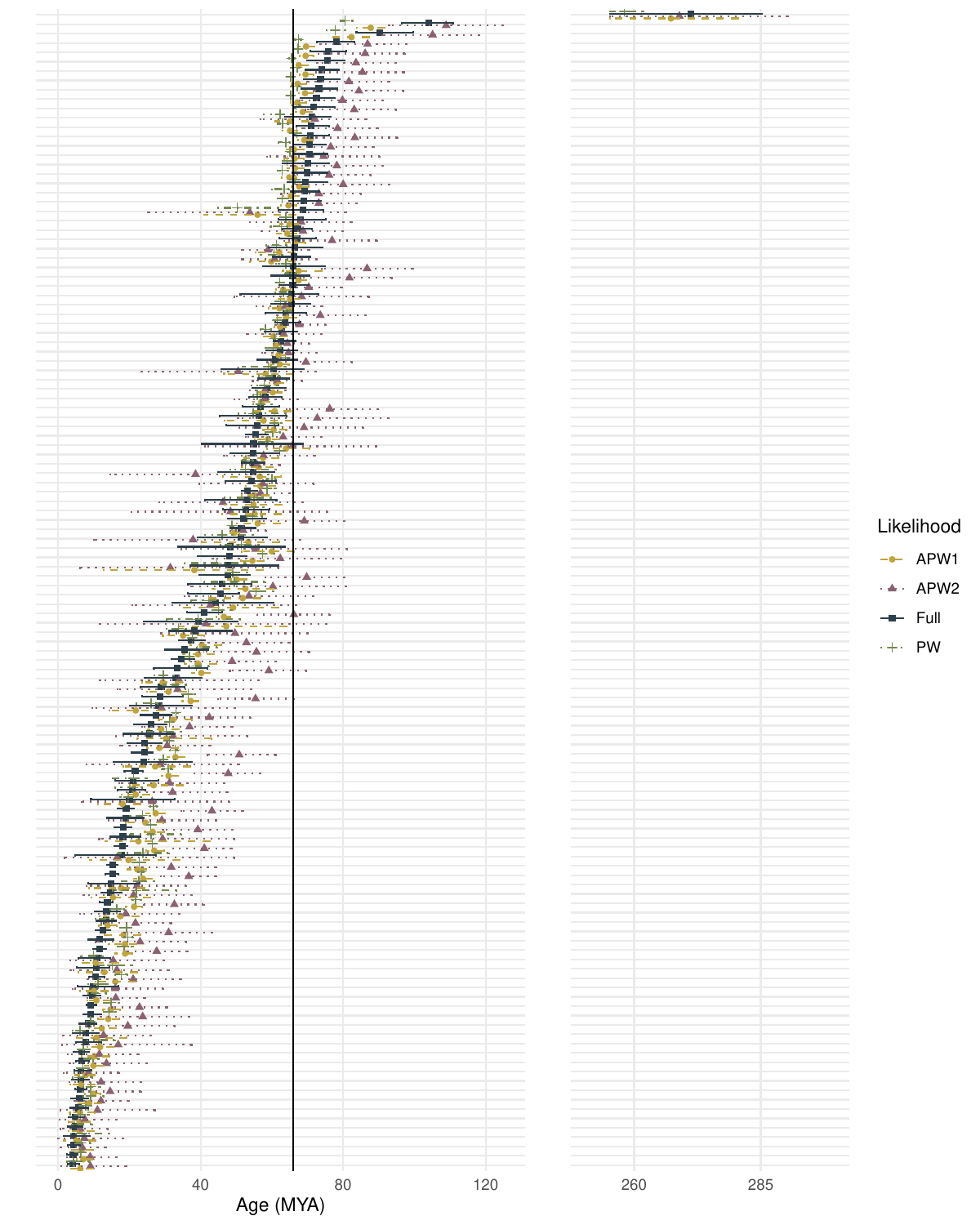}
  \caption{Node age estimates for all nodes of bird tree topology. The APW1, APW2, Full, and PW
  likelihood-based point estimates are indicated by the yellow circles, red triangles, black squares,
  and gree crosses, respectively. The corresponding 95\% credible intervals are given by the 
  yellow dashed, green dotted, black solid, and green dot-dashed lines, respectively. } 
  \label{fig:all_nodes}
\end{figure}

\begin{figure}[ht!]
  \centering
  \includegraphics[angle=0,width=0.9\textwidth]{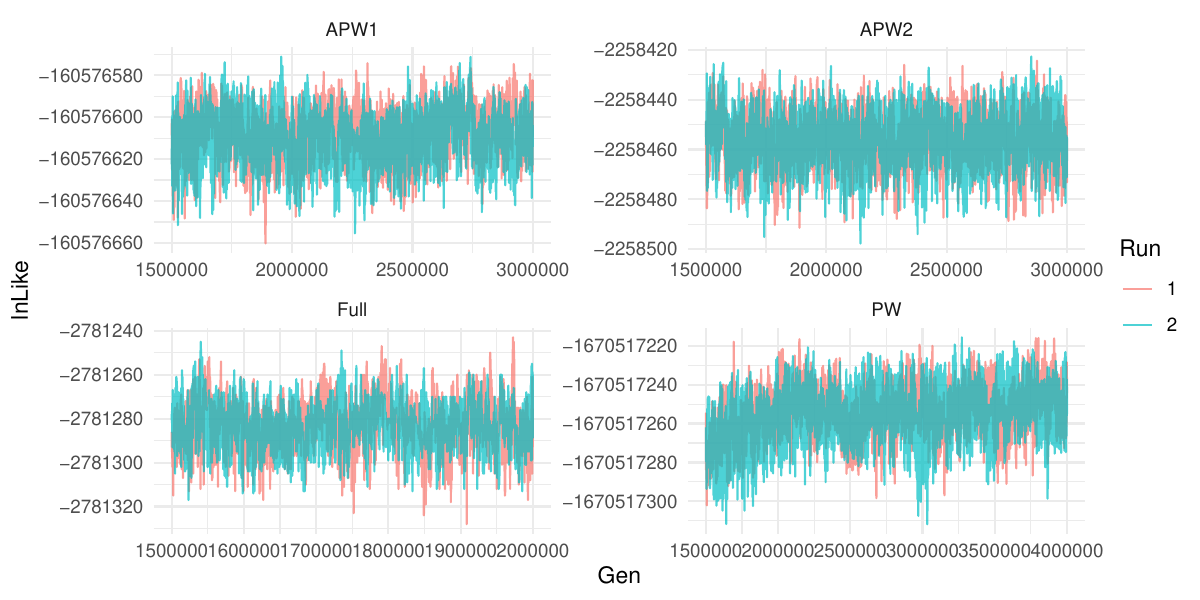}
  \caption{Trace plots showing log likelihood values for two independent runs per likelihood. 
  } 
  \label{fig:mcmc_trace}
\end{figure}

\end{document}